\documentclass[twocolumn,english,prb]{revtex4}
\usepackage[T1]{fontenc}
\usepackage[latin9]{inputenc}
\usepackage{amsmath}
\usepackage{graphicx}
\usepackage{amssymb}

\makeatletter

\newcommand{\lyxdot}{.}

\@ifundefined{textcolor}{}
{%
 \definecolor{BLACK}{gray}{0}
 \definecolor{WHITE}{gray}{1}
 \definecolor{RED}{rgb}{1,0,0}
 \definecolor{GREEN}{rgb}{0,1,0}
 \definecolor{BLUE}{rgb}{0,0,1}
 \definecolor{CYAN}{cmyk}{1,0,0,0}
 \definecolor{MAGENTA}{cmyk}{0,1,0,0}
 \definecolor{YELLOW}{cmyk}{0,0,1,0}
 }

\makeatother
\makeatother

\usepackage{babel}

\begin{document}

\title{The  classical and quantum dynamics of the inhomogeneous Dicke model and its  Ehrenfest time }


\author{Oleksandr Tsyplyatyev }

\affiliation{Department of Physics, University of Basel, Klingelbergstrasse 82,
CH-4056 Basel, Switzerland }

\author{Daniel Loss}

\affiliation{Department of Physics, University of Basel, Klingelbergstrasse 82,
CH-4056 Basel, Switzerland }

\date{\today}
\begin{abstract}
We show that in the few-excitation regime the classical and quantum time-evolution
of the inhomogeneous Dicke model for $N$ two-level
systems coupled to a single boson mode agree for $N\gg1$. In the presence
of a single excitation only,
the leading term in an $1/N$-expansion of the classical equations of
motion reproduces the result of the Schr\"odinger equation. For a small
number of excitations, the numerical solutions of the classical and quantum
problems become equal for $N$ sufficiently large. By solving
the Schr\"odinger equation exactly for two excitations and a particular
inhomogeneity we obtain $1/N$-corrections which lead to a significant
difference between the classical and quantum solutions at a new time
scale which we identify as an Ehrenferst time, given by $\tau_{E}=\sqrt{N/\left\langle g^{2}\right\rangle }$,
where $\sqrt{\left\langle g^{2}\right\rangle }$ is an effective coupling strength between
the two-level systems and the boson.
\end{abstract}
\maketitle

\section{Introduction}

The recent experimental advances on cold atoms in optical cavities
\cite{Esslinger}, Bose-Einstein condensation of exciton polaritons
\cite{Littlewood}, and observation of vacuum Rabi oscillations \cite{Weisbuch}
in semiconductor microcavities renewed interest in light-matter
interaction in the quantum coherent regime. These studies were motivated
by an observation made by Dicke \cite{Dicke} long ago who realized that radiation
from $N$ identical two-level systems (spins 1/2) cannot be treated
as a sum of $N$ independent radiative processes but rather as a collective
quantum phenomenon that involves all $N$ spins and a photon mode
even on the level of perturbation theory. Also, several schemes based
on light-matter interaction to couple spatially separated spins that
had been originally proposed as an element of a quantum computing
device \cite{CiracZoller,Burkard,Lukin,Trif} were recently improved
by a suggestion to use qubits constructed out of many spins to enhance
coupling with the optical mode \cite{Imamoglu08} due to the superradiant
effect.

For instance, considerable attention was paid experimantally to the
$\sqrt{N}$-enhancement of the light-matter coupling \cite{Esslinger,Rempe}.
In typical set-ups the spins are spatially separated, therefore the
excitation energies of different spins may be different as they are
affected by local forces that typically vary across the sample. The
coupling strength to the light mode also varies as different spins are
located at different positions of the mode due to a different amplitude
of the electromagnetic field. Understanding of such inhomogeneities
is important to find the practical limitations on the decoherence
time of the system when, for instance, one designs a quantum computing
device \cite{CiracZollerKimbleMabuchi,Burkard,Trif}. Also, the inhomogeneities
are unavoidable and should be important in a system like a semiconductor
quantum dot optical amplifier or laser \cite{Deppe,Forchel,Nakata}.

On the theoretical side, the homogeneous Dicke model, which describes a bath of
$N$ equivalent spins-1/2 with excitation (Zeeman) energy $\epsilon$ coupled to
a quantized bosonic mode $\omega$ with the same coupling constants
$g$, was diagonalized exactly in Ref. \onlinecite{TavisCummings}. The influence
of inhomogeneities of the coupling constants $g_{j}$ and Zeeman energies
$\epsilon_{j}$ on the single excitation dynamics was analyzed exactly
in Refs. \onlinecite{Cummings,OT_Loss}. It was shown that the boson occupation oscillates
in time with a single Rabi frequency $\Omega=\sqrt{N\left\langle g^{2}\right\rangle }$,
where $\sqrt{\left\langle g^{2}\right\rangle }$ is an effective coupling
when only the coupling constants $g_{j}$ are inhomogeneous but with constant Zeeman
energies.  If the
Zeeman energies $\epsilon_{j}$ are also inhomogeneous but spread narrower than the threshold
given by $\Omega$ this single frequency acquires a small Lamb-like
shift, whereas for a spread exceeding $\Omega$ the boson decays completely
in time. 

In this paper we show that the solution to the classical Hamilton
equations of motion matches the solution of the time-dependent Schr\"odinger
equation when the number of spins is large, i.e. $N\gg1$, while the number
of excitations $p$ is still small, i.e. $p\ll N$. For a single excitation ($p=1$) the
leading order in an $1/N$-expansion of the classical equations agrees
with the quantum one. For a few excitations such correspondence does
not hold, but for $p=2,3$ the numerical solutions of both
classical equations of motion and Schr\"odinger equation agree for $N\gg1$. 
It is plausible to assume that in leading
$1/N$-order the same correspondence holds for $p>3$. The numerical
treatment of the Schr\"odinger equation with a large number of spins
is possible since the Fock space scales only as a power of $N$ ($N^{2},\, N^{3},\dots$)
in the few-excitation subspaces.

As the classical equations of motion for $p>1$ can also be mapped
on the Schr\"odinger equation in the single excitation subspace in
leading $1/N$-order the already available quantum result can be used
to analyze the classical equations of motion for few excitations ($p\ll N$).
For $p$ excitations with $p>1$ we obtain the dynamics by simply
rescaling the solution derived in \cite{OT_Loss} by $p$. This extends
the single-excitation quantum solution to the case of few excitations
when $N\gg1$.

To assess the validity of the classical approximation for $p>1$ excitations
we solve the Schr\"odinger equation exactly in the two-excitation subspace
with inhomogeneity in the coupling constants only and compare it with
the classical solution. When $N$ is small both solutions are completely
different. For large $N$ we perform an $1/N$-expansion of the
quantum solution and recover the classical result in leading order.
Subleading $1/N$-corrections cause deviations between quantum and
classical dynamics which become significant at a large time scale
$\tau_{E}=\sqrt{N/\left\langle g^{2}\right\rangle }$ for $p\ll N$.
We refer to this time scale as an Ehrenfest time, defined here as
the time where the quantum dynamics starts to differ from the classical
dynamics.

Also, having found a quantum solution for $p=2$ we study it separately
and in particular compare it with the $p=1$ quantum dynamics. We
find that inhomogeneity of the coupling constants results in a different
spectrum when $N$ is finite: in the subspace with $p=1$ there is
only one harmonic mode with a single frequency in the time-dependent
occupation number of the boson, and for $p=2$ there
are $N$ discrete harmonic modes that form a continuum spectrum
in the limit of large $N$. Such a mechanism can lead to destructive
interference, thus to decay, of the excitations caused solely by
the inhomogeneity of the coupling constants when $p>1$. But, as pointed
out already, for $p=2$ we find that the leading $1/N$-term recovers
the single frequency dynamics in accordance with the classical solution.
The decay due to inhomogeneous coupling constants thus manifests
itself only in the first subleading $1/N$-correction. We find that
this contribution is an oscillatory mode with frequency $\frac{3}{2}\Omega$
and a slowly decaying envelope. The decay behavior is essentially non-exponential
with a long power-law tail and the decay time is $\tau_{g}\sim\sqrt{N/\left\langle g^{2}\right\rangle }$,
where $\sqrt{\left\langle g^{2}\right\rangle }$ is a characteristic
coupling. This decay occurs on the same time scale as the Ehrenfest
time $\tau_{E}$ defined above. Thus, it can be described correctly only by the
Schr\"odinger equation (and not by the classical one).

In our theoretical analysis we assume the following ideal experiment.
The spin bath is prepared in the ground state, e.g. dynamically or
by the thermal cooling. The non-equilibrium dynamics of the boson
is then initialized by a short radiation pulse from an external source
which populates the boson mode with a few excitations like in \cite{Solano,Gross,Schwendimann}.
The dissipation of the boson mode, e.g. leakage of the photons through
the mirrors that define an optical cavity can be used to detect the
dynamics, similarly to the measurements performed on semiconductor
quantum well microcavities\cite{Weisbuch,Yamamoto,Littlewood}, for
the limiting case where the cavity leakage time exceeds the internal
time scale. 

The rest of the paper is organized as follows. In Section II we discuss
general properties of the inhomogeneous Dicke model. In Section III
we quote the already known solution to the Schr\"odinger equation in
the single excitation subspace. In Section IV we construct the classical
analog of the inhomogeneous Dicke model. Section V contains the exact
solution of the Schr\"odinger equation in the two-excitation subspace
for the inhomogeneous couplings only. In Section VI we compare the
numerical solution of the classical and the quantum equations of motions
for two and three excitations in the limit of many spins. Section
VII contains a discussion of applicability of the classical approximation.
In the Appendix  we give some details on the calculation of the $1/N$-correction.

\section{Inhomogeneous Dicke model}

The Hamiltonian for the Dicke model that describes the interaction
between a set of $N$ spins 1/2 with excitation energies $\epsilon_{j}$
and a single bosonic mode of frequency $\omega$ is given by

\begin{equation}
H=\omega b^{\dagger}b+\sum_{j=1}^{N}\epsilon_{j}S_{j}^{z}+\sum_{j=1}^{N}g_{j}\left(S_{j}^{+}b+S_{j}^{-}b^{\dagger}\right),\label{eq:Dicke_model}\end{equation}
where $S_{j}^{\pm}=S_{j}^{x}\pm iS_{j}^{y}$, $S_{j}^{z}$ are spin
1/2 operators, $b\left(b^{\dagger}\right)$ are the standard Bose
annihilation (creation) operators. The coupling constants $g_{j}$
are typically given as dipole matrix elements and thus are, in general,
complex numbers. Since their phases can be eliminated by a
unitary transformation, we treat $g_{j}$ as real and positive numbers. 

In the present paper we assume that the boson mode is tuned in resonance
with the spins $\left\langle \epsilon_{j}\right\rangle =\omega$,
where $\left\langle \dots\right\rangle =\sum_{j}\dots/N$. If the
boson mode is strongly detuned, $\left|\left\langle \epsilon_{j}\right\rangle -\omega\right|\gg\sqrt{\left\langle g^{2}\right\rangle }$,
the interaction between them is weak and the model Eq. (\ref{eq:Dicke_model})
can be analyzed perturbatively \cite{CiracZollerKimbleMabuchi}. Also
note that the inhomogeneities of $g_{j}$ and/or $\epsilon_{j}$
forbids to represent the Hamiltonian Eq. (\ref{eq:Dicke_model}) in
terms of the total angular momentum operators $J_{\alpha}=\sum_{j}S_{j}^{\alpha}$, ${\alpha}=x,y,z$.
. 

The total number of spin-boson excitations, $L=n+\sum_{j}S_{j}^{z}$,
is conserved by the model Eq. (\ref{eq:Dicke_model}), where $n=b^{\dagger}b$
is the bosonic occupation number. The eigenvalue $c$ of $L$ labels
the subspace of the Hamiltonian with a given total number of excitations.

We restrict ourselves to a small number of excitations, $p\ll N$.
In the following we assume that the spins can be prepared in the ground
state with each spin in its low Zeeman state. The bosonic mode is
assumed to be occupied by $p$ bosons initially, the time evolution
is restricted to the subspace with $c=-N/2+p$. Then the leakage of
the boson mode to the outside world can be used to monitor the time
dynamics of the system by detecting the leaked mode at given subsequent
instances in time.

\section{Single excitation}

The time dynamics of Eq. (\ref{eq:Dicke_model}) for a single excitation
was analyzed in detail in Ref. \onlinecite{OT_Loss}. Here we only
quote the explicit form of the corresponding Schr\"odinger equation
and the main results derived from it.

The time evolution is restricted the the subspace with $c=-N/2+1$
and is described by the general state

\begin{equation}
\left|\Psi\left(t\right)\right\rangle =\alpha\left(t\right)\left|\Downarrow,1\right\rangle +\sum_{j=1}^{N}\beta_{j}\left(t\right)\left|\Downarrow\uparrow_{j},0\right\rangle ,\label{eq:psi_1}\end{equation}
where $\alpha\left(t\right)$ and $\beta_{j}\left(t\right)$ are normalized
amplitudes, $\left|\alpha\left(t\right)\right|^{2}+\sum_{j}\left|\beta_{j}\left(t\right)\right|^{2}=1$,
of finding either a state with one boson and no spin excitations present
or a state with no boson and the $j^{th}$-spin excited (flipped).
As initial condition we will assume throughout  (with one exception discussed at the end) that initially only bosonic excitations are present while each spin is in its individual ground state, i.e. $\alpha\left(t=0\right)=1$. The state $\left|\Psi\left(t\right)\right\rangle $ from Eq. (\ref{eq:psi_1})
describes then the time evolution of an initial product state $\left|\Downarrow,1\right\rangle$ into 
an entangled state  formed by a coherent superposition of N+1 states, where each $\left|\Downarrow\uparrow_{j},0\right\rangle$ contains  an excited spin and no boson. This entangled state can be viewed as a (para-) magnon state in the uniform limit.
In other words, the initial bosonic excitation gets coherently spread out over the entire system in course of time.

Inserting $\left|\Psi\left(t\right)\right\rangle $ from Eq. (\ref{eq:psi_1})
into the time-dependent Schr\"odinger equation we get \begin{eqnarray}
-i\dot{\alpha}\left(t\right) & = & \sum_{j}g_{j}\beta_{j}\left(t\right),\label{eq:eqs_of_motion_1}\\
-i\dot{\beta}_{k}\left(t\right) & = & \left(\epsilon_{j}-\omega\right)\beta_{k}\left(t\right)+g_{k}\alpha\left(t\right).\nonumber \end{eqnarray}
This set of coupled equations can be solved explicitly via Laplace
transformation. We use the same approach to solve the Schr\"odinger
equation in the two excitation subspace in Section V of this paper.

If the number of spins is large, $N\gg1$, the sum over $j$ in the
exact solution of Eq. (\ref{eq:eqs_of_motion_1}) can be substituted
by an integral. In this continuum limit the discrete set of $\epsilon_{j}$
and $g_{j}$ become continuous variables characterized by distribution
functions $Q\left(g\right)$ and $P\left(\epsilon\right)$. Any distribution
function of $g$ results only in a renormalized coupling constant
$\sqrt{\left\langle g^{2}\right\rangle }$ and the dynamics of the
boson is not affected in any other way.

Different distribution functions of $\epsilon$ result in qualitatively
different regimes of the dynamics. Let us choose $P\left(\epsilon\right)$
as a rectangular pulse shape of width $\Delta$ centered around $\omega$,
\begin{equation}
P\left(\epsilon\right)=\theta\left(-\epsilon+\omega+\Delta/2\right)\theta\left(\epsilon-\omega+\Delta/2\right),\label{eq:P_eps}\end{equation}
where $\theta(x)$ is  the Heaviside step-function.
It was shown that when the inhomogeneity is below a certain threshold,
$\Delta/\Omega\ll1$, where $\Omega=\sqrt{N\left\langle g^{2}\right\rangle }$
is the collective Rabi frequency, the boson excitation, $\left\langle n\right\rangle =\left|\alpha\left(t\right)\right|^{2}$,
does not decay, i. e. \begin{equation}
\left\langle n(t)\right\rangle =\cos^{2}\left(\Omega t\right).\label{eq:1_excitation_1_frequency}\end{equation}
The corrections to this result are small and are on the order of $\Delta/\Omega$.
In the opposite limit, $\Delta/\Omega\gg1$, the spins act as the
thermal bath at zero temperature. The bosonic excitation decays completely
and exponentially, \begin{equation}
\left\langle n(t)\right\rangle =\exp\left(-t/t_{2}\right),\label{eq:1_excitation_exp}\end{equation}
with the decay time $t_{2}=2\Delta/\pi\Omega^{2}$. In the intermediate
regime, $\Delta\simeq\Omega$, the decay is partial and the decay
law is a combination of exponential and inverse-power laws.

\section{Classical analogy}

Here we construct a classical version of the inhomogeneous Dicke model.
Using Dirac's analogy \cite{Sakurai} we change the boson operator
$b$ in the model Eq. (\ref{eq:Dicke_model}) to a classical complex variable
$a=a_{x}+ia_{y}$, and the spin operators $\mathbf{S}_{j}$ to a set
of $N$ vectors $\mathbf{C}_{j}=\left(C_{j}^{x},C_{j}^{y},C_{j}^{z}\right)$
of length $\left|\mathbf{C}_{j}\right|=1/2$. These classical degrees
of freedom obey the Poisson bracket relations which are obtained from the bosonic
and spin commutation relations via the ansatz $\left[,\right]\rightarrow-i\left[,\right]_{cl}$:
$\left[C_{\alpha},C_{\beta}\right]_{cl}=-\epsilon_{\alpha\beta\gamma}C_{\gamma}$,
and $\left[a,a^{*}\right]_{cl}=i$.

The Hamilton equation of motion for the $j^{\textrm{th}}$ spin, $\dot{\mathbf{C}}_{j}=\left[H,\mathbf{C}_{j}\right]_{cl}$,
is a Bloch equation \begin{equation}
\dot{\mathbf{C}}_{j}=\mathbf{B}_{j}\times\mathbf{C}_{j},\label{eq:C_j}\end{equation}
where the in-plane component of the effective magnetic field is the
complex bosonic field, and the perpendicular component is the single
spin excitation energy, $\mathbf{B}_{j}=\left(2g_{j}a_{x},2g_{j}a_{y},\epsilon_{j}-\omega\right)$.
The Hamilton equation of motion for $a$ is a feedback to the bosonic
field from the in-plane component of all spins, \begin{equation}
\dot{a}=-i\sum_{j}g_{j}C_{j}^{-},\label{eq:a}\end{equation}
where $C_{j}^{-}=C_{j}^{x}-iC_{j}^{y}$. Generally, these differential
equations can be solved numerically with the initial conditions $\mathbf{C}_{j}\left(0\right)=\left(0,0,-1/2\right)$
and $a\left(0\right)=\sqrt{p}$ to obtain the time-dependent solution
$\mathbf{C}_{j}\left(t\right)$ and $a\left(t\right)$ explicitly.
The time-dependent value of the bosonic field is $n\left(t\right)=\left|a\left(t\right)\right|^{2}$.

For a small number of excitations, $p\ll N$, Eq. (\ref{eq:C_j})
simplifies. The quantity $L=\left|a\right|^{2}+\sum_{j}C_{j}^{z}=-N/2+p$
is conserved during the evolution governed by Eqs. (\ref{eq:C_j},
\ref{eq:a}). Thus, at any instance of time $\sum_{j}C_{j}^{z}\approx-N/2$,
i. e. if the dynamics starts with only a few bosonic excitations,
the spins cannot 'flip' during the evolution. Using the approximation $C_{j}^{z}\left(t\right)\approx-1/2$,
the equation for $C_{j}^{z}\left(t\right)$ drops out from Eq. (\ref{eq:C_j})
and the remaining two equations are 

\begin{equation}
\dot{C}_{j}^{-}=-i\left(\epsilon_{j}-\omega\right)C_{j}^{-}-ig_{j}a.\label{eq:Cm_j}\end{equation}

When $p=1$, the Hamilton Eqs. (\ref{eq:a}, \ref{eq:Cm_j}) with
the initial conditions above coincide formally with the Schr\"odinger
Eq. (\ref{eq:eqs_of_motion_1}). By direct comparison we can establish
the correspondence between the quantum mechanical amplitudes and the
classical variables: the classical field $a$ is the amplitude $\alpha$
and the in-plane component of the spin vector $C_{j}^{-}$ is the
amplitude $\beta_{j}$. Note that the classical spins are not averages
of the spin operators, $\left\langle \mathbf{S}_{j}\right\rangle \equiv0$,
but instead they are connected with the quantum mechanical amplitudes. The
solution of the dynamical Eqs. (\ref{eq:a}, \ref{eq:Cm_j}) is the
same as the solution of Eq. (\ref{eq:eqs_of_motion_1}). 

When the number of excitations at the initial time is $p>1$, but
still $p\ll N$, the mapping of the classical equations on Eq. (\ref{eq:eqs_of_motion_1}),
approximation (\ref{eq:Cm_j}), still holds but the initial condition
for Eq. (\ref{eq:eqs_of_motion_1}) is different: $\alpha\left(0\right)=\sqrt{p}$.
This difference results then only in a renormalization of $\left\langle n(t)\right\rangle $ [obtained
from Eq. (\ref{eq:eqs_of_motion_1})] by 
$p$. 

When the number of excitations is large $p\geq N$, the z-components of
the classical spins deviate significantly from their initial values
during the evolution and the approximation (\ref{eq:Cm_j}) is not
valid. In this regime the classical Eqs. (\ref{eq:C_j}, \ref{eq:a})
have to be solved numerically.

\section{Two-excitation regime and inhomogeneous coupling constants}

In this section we consider the dynamics of two excitations for a system with inhomogeneous
coupling constants  $g_{j}$ but constant Zeeman energies $\epsilon_j=\omega$.

The time evolution of the two excitations is restricted to the subspace
with $c=-N/2+2$ and is described by the general state \begin{equation}
\left|\Psi\left(t\right)\right\rangle=\alpha\left(t\right)\left|2,\Downarrow\right\rangle +\sum_{j}\beta_{j}\left|1,\Downarrow\uparrow_{j}\right\rangle +\sum_{i>j}\gamma_{ij}\left|0,\Downarrow\uparrow_{i}\uparrow_{j}\right\rangle ,\label{eq:wave_function}\end{equation}
where $\alpha\left(t\right)$, $\beta_{j}\left(t\right)$, and $\gamma_{ij}\left(t\right)$
are the normalized amplitudes, $\left|\alpha\left(t\right)\right|^{2}+\sum_{j}\left|\beta_{j}\left(t\right)\right|^{2}+\sum_{i>j}\left|\gamma_{ij}\right|^{2}=1$,
of the state with two bosonic excitations, a state with one bosonic excitation
and the $j^{th}$ spin excited, and a state with no bosonic excitation
and the $i^{th}$ and $j^{th}$ spins excited (with $i\neq j$). The amplitude $\gamma_{ij}$
is defined such that $\gamma_{ij}=0$ if $j\geq i$.

The conservation law can be used to simplify the Hamiltonian Eq. (\ref{eq:Dicke_model}).
We subtract $\omega L$  from Eq. (\ref{eq:Dicke_model}), which only changes an irrelevant overall phase of $\left|\Psi\left(t\right)\right\rangle $, to eliminate
the first two terms. Note that the second term will not be zero away
from the resonance $\omega\neq\epsilon$. Inserting $\left|\Psi\left(t\right)\right\rangle $
into the time-dependent Schr\"odinger equation we then obtain,

\begin{eqnarray}
-i\dot{\alpha}\left(t\right) & = & \sqrt{2}\sum_{j}g_{j}\beta_{j}\left(t\right),\label{eq:dynamic_equations_t}\\
-i\dot{\beta}_{k}\left(t\right) & = & \sqrt{2}g_{k}\alpha\left(t\right)+\sum_{j<k}g_{j}\gamma_{kj}\left(t\right)+\sum_{j>k}g_{j}\gamma_{jk}\left(t\right),\nonumber \\
-i\dot{\gamma}_{kl}\left(t\right) & = & \left(g_{k}\beta_{l}\left(t\right)+g_{l}\beta_{k}\left(t\right)\right)\left(1-\delta_{kl}\right).\nonumber \end{eqnarray}
The initial condition, $\alpha\left(0\right)=1$ and $\beta_{j}\left(0\right)=\gamma_{ij}\left(0\right)=0$,
which we further assume corresponds to the doubly occupied boson mode
at the initial time. The physical observable of interest is the
time-dependent value of the boson occupation number $\left\langle n \left(t\right)\right\rangle $,
which can be expressed in terms of the amplitudes $\alpha\left(t\right)$
and $\beta_{j}\left(t\right)$ as \begin{equation}
\left\langle n \left(t\right)\right\rangle =2\left|\alpha\left(t\right)\right|^{2}+\sum_{j}\left|\beta_{j}\left(t\right)\right|^{2},\label{eq:n_ab}\end{equation}
where $\left\langle \dots\right\rangle=\left\langle \Psi\left(t\right)\middle|\dots\middle|\Psi\left(t\right)\right\rangle $
is the time-dependent expectation value.

\subsection{General solution}

We use the Laplace transform, $A\left(s\right)=\int_{0}^{\infty}dtA\left(t\right)e^{st}$,
to solve the set of equations, Eq. (\ref{eq:dynamic_equations_t}).
In the Laplace domain Eq. (\ref{eq:dynamic_equations_t}) is a set
of linear algebraic equations, \begin{equation}
\begin{array}{rcl}
-i\left(s\alpha-1\right) & = & \sqrt{2}\sum_{j}g_{j}\beta_{j}\\
-is\beta_{k} & = & \sqrt{2}g_{k}\alpha+\sum_{j<k}g_{j}\gamma_{kj}+\sum_{j>k}g_{j}\gamma_{jk}\\
-is\gamma_{kl} & = & \left(g_{k}\beta_{l}+g_{l}\beta_{k}\right)\left(1-\delta_{kl}\right),\end{array}\label{eq:dynamic_equations_s}\end{equation}
that can be explicitly solved. The substitution of $\alpha\left(s\right)$
and $\gamma_{kl}\left(s\right)$ as functions of $\beta_{k}\left(s\right)$,
that are obtained from the first and the last lines, into the middle
line gives the following set of equations for $\beta_{k}$ only,
\begin{equation}
\left(-s^{2}+2g_{k}^{2}-\sum_{j}g_{j}^{2}\right)\frac{\beta_{k}}{g_{k}}=3\sum_{j}g_{j}\beta_{j}-i\sqrt{2}.\end{equation}
Each $\beta_{k}\left(s\right)$ is easily found from the the above
equation since $\sum_{j}g_{j}\beta_{j}\left(s\right)$ on the right
hand side is the same in each equation for all $\beta_{k}\left(s\right)$.
Then the sum is found self-consistently and we obtain the solution
for the amplitude as \begin{equation}
\beta_{k}\left(s\right)=\frac{-i\sqrt{2}g_{k}}{-s^{2}+2g_{k}^{2}-N\left\langle g^{2}\right\rangle }\frac{1}{1-3\left\langle \frac{g_{j}^{2}N}{-s^{2}+2g_{j}^{2}-N\left\langle g^{2}\right\rangle }\right\rangle },\label{eq:bs_k}\end{equation}
where the average is the sum over all spins $\left\langle \dots\right\rangle =\left(\sum_{j}\dots\right)/N$.
The other two amplitudes are found from the first and the third lines
of Eq. (\ref{eq:dynamic_equations_s}) by substitution of the above
solution for $\beta_{k}\left(s\right)$, \begin{equation}
\alpha\left(s\right)=\frac{1}{s}\frac{1-\left\langle \frac{g_{j}^{2}N}{-s^{2}+2g_{j}^{2}-N\left\langle g^{2}\right\rangle }\right\rangle }{1-3\left\langle \frac{g_{j}^{2}N}{-s^{2}+2g_{j}^{2}-N\left\langle g^{2}\right\rangle }\right\rangle },\label{eq:a_s}\end{equation}
 \begin{equation}
\gamma_{kl}\left(s\right)=\frac{i}{s}\left(g_{k}\beta_{l}+g_{l}\beta_{k}\right)\left(1-\delta_{kl}\right).\end{equation}

The main focus of our interest will be on Eqs. (\ref{eq:bs_k}, \ref{eq:a_s})
as the observable quantity $\left\langle n \left(t\right)\right\rangle $
depends only on $\alpha\left(t\right)$ and $\beta_{k}\left(t\right)$.
These time-dependent amplitudes can be obtained from Eqs. (\ref{eq:bs_k},
\ref{eq:a_s}) by the inverse Laplace transform. The analytic structure
of Eqs. (\ref{eq:bs_k}, \ref{eq:a_s}) is governed, in general, by
a set of poles given by the roots of denominators which depend on
a particular set of $g_{j}$. For instance, if the number of spins
$N$ is small there are $2N$ conjugated complex roots. The inverse
Laplace transforms of $\alpha\left(s\right)$ and $\beta_{k}\left(s\right)$
will be a sum of $N$ discrete harmonic modes in contrast to the single
excitation dynamics where in such a setup there is just a single pair
of roots independent of the particular set of $g_{j}$, see Section
III, and there is only a single harmonic mode in the dynamics of the
boson occupation number, see Eq. (\ref{eq:1_excitation_1_frequency}).
Such a result marks a qualitative difference in the dynamics of the
single- and two-excitation subspaces.

\subsection{ Time-evolution in the continuum limit of many spins }

In this section we study the limit of many spins, i.e. $N\gg 1$. The sum over $j$
in Eqs. (\ref{eq:bs_k}, \ref{eq:a_s}) can be substituted by an integral
over a distribution function of $g$, $\sum_{j}\dots\rightarrow N\int_{0}^{\infty}dgQ\left(g\right)\dots$
In the continuum limit some poles can merge together, forming branch
cuts, and some poles can separate themselves from the others. The
inverse Laplace transform of the branch cuts will become a decay function
in the time domain and the separate poles will contribute  a set
of harmonic modes.

The analytic structure of $\alpha\left(s\right)$ and $\beta_{k}\left(s\right)$
explicitly depends on the particular form of $Q\left(g\right)$.

\subsection{Uniform distribution function}

\begin{figure}[pt]
\centering\includegraphics[width=0.9\columnwidth]{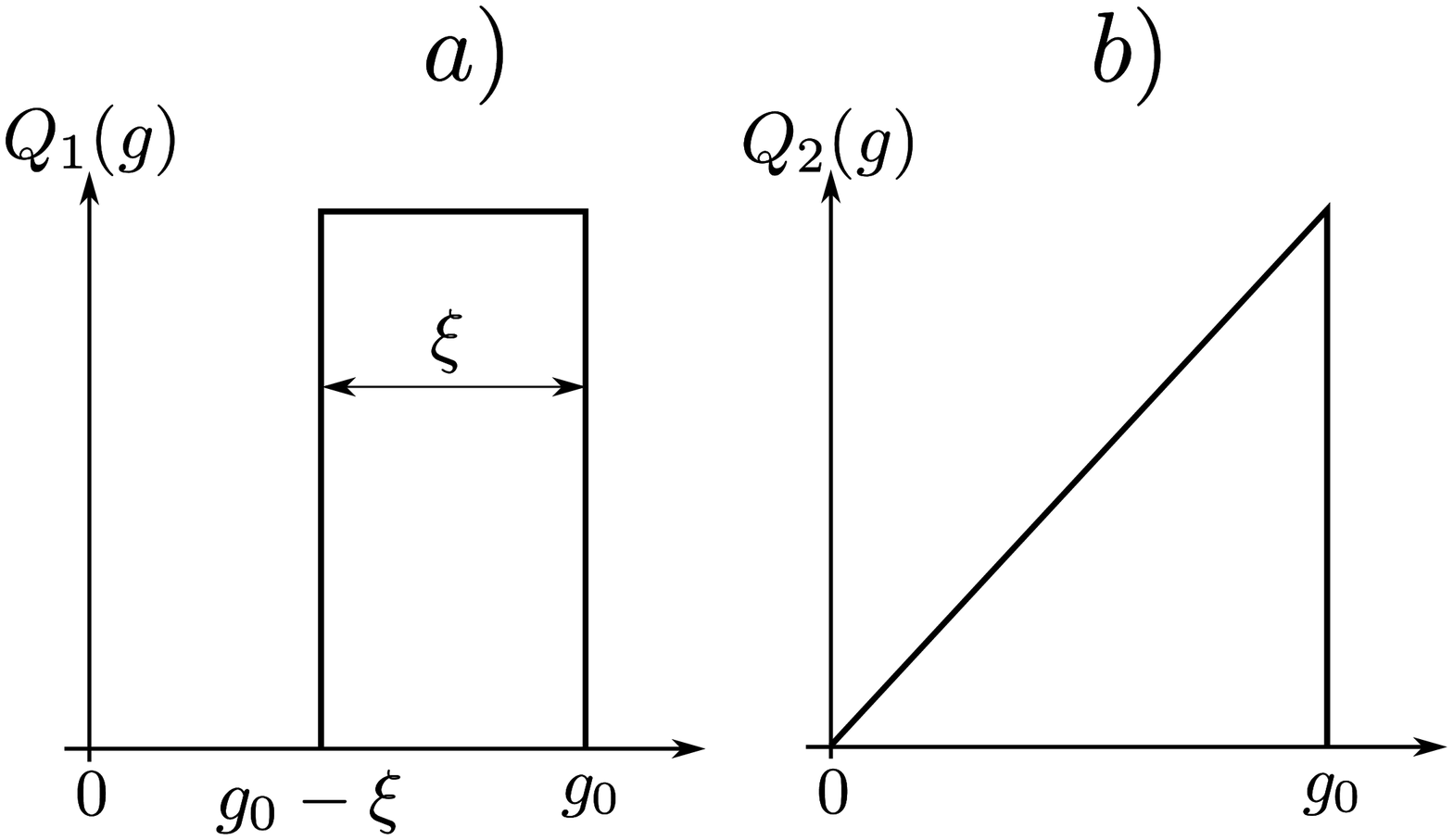} 

\caption{Distribution functions of $g$ that are used to evaluate the sums in Eqs.
(\ref{eq:bs_k}, \ref{eq:a_s}). (a) The uniform distribution function
$Q_{1}\left(g\right)$ has a maximum coupling strength $g_{0}$ and a width $\xi$
which can vary from $0$ (i.e. homogeneous coupling constants) to $g_{0}$ (i.e. maximally
inhomogeneous coupling constants). (b) The sawtooth distribution function $Q_{2}\left(g\right)$
describes a non-uniform spread of the coupling constants $g_j$ from $0$ to $g_{0}$.\label{fig:P_gs}}

\end{figure}

To be specific we consider a set of coupling constants which are uniformly
distributed from a minimum value $g=g_{0}-\xi$ to a maximum value
$g=g_{0}$, \begin{equation}
Q_{1}\left(g\right)=\theta\left(-g+g_{0}\right)\theta\left(g-g_{0}+\xi\right)/\xi . 
\label{eq:Q1_g}\end{equation}
The coupling
constants cannot be negative, so $\xi$ can vary from $\xi=0$ (e.
g. all couplings are the same and are equal to $g_{0}$) to $\xi=g_{0}$
(e. g. the couplings are evenly distributed from 0 to $g_{0}$), see
Fig. \ref{fig:P_gs}a. A useful property of this distribution function $Q_{1}$
is that a small and a large inhomogeneity can be analyzed on the same footing.
Another distribution function will be considered in the next subsection. 

Turning the sum in Eqs. (\ref{eq:bs_k}, \ref{eq:a_s}) into an integral
and using $Q_{1}\left(g\right)$, we obtain\begin{widetext}\begin{equation}
\left\langle \frac{g_{j}^{2}N}{-s^{2}+2g_{j}^{2}-N\left\langle g^{2}\right\rangle }\right\rangle =-\frac{N}{2}\left(\frac{\sqrt{-s^{2}-N\left\langle g^{2}\right\rangle }}{\sqrt{2}\xi}\arctan\left(\frac{\sqrt{2}\xi\sqrt{-s^{2}-N\left\langle g^{2}\right\rangle }}{s^{2}+N\left\langle g^{2}\right\rangle -2g_{0}\left(g_{0}-\xi\right)}\right)+1\right)\label{eq:sum_even_couplings}, \end{equation}
 \end{widetext}where $\left\langle g^{2}\right\rangle =g_{0}^{2}-\xi g_{0}+\xi^{2}/3$.%

\begin{figure}[pt]
\centering\includegraphics[width=0.9\columnwidth]{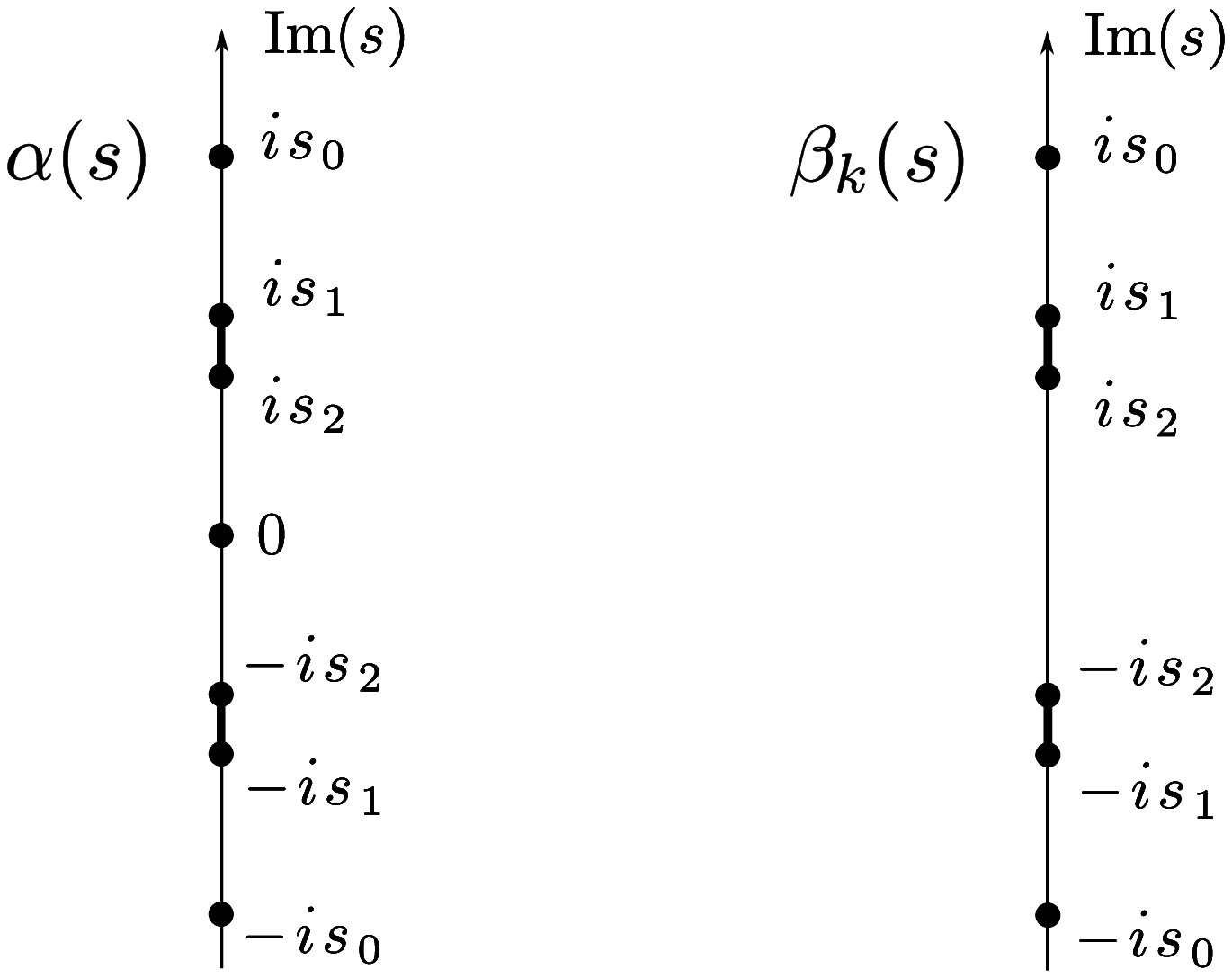} 

\caption{\label{fig:analytic_structure}Analytic structure of the time-dependent
quantum mechanical amplitudes for $p=2$ excitations in the Laplace
domain, Eqs. (\ref{eq:bs_k}, \ref{eq:a_s}), in the continuum approximation
calculated using the distribution functions $Q_{1}\left(g\right)$ and
$Q_{2}\left(g\right)$. Separated dots, $0,\pm is_{0}$, are poles
and the dots, $\pm is_{1,2}$, connected by bold lines, are branch points.
The bold lines are the corresponding branch cuts.}

\end{figure}

The analytic structure of $\alpha\left(s\right)$ from Eq. (\ref{eq:a_s})
with the sum from Eq. (\ref{eq:sum_even_couplings}) is the following.
There are three poles and two branch cuts, see Fig. \ref{fig:analytic_structure}.
Thus, the inverse Laplace transform has two contributions $\alpha\left(t\right)=\alpha_{p}\left(t\right)+\alpha_{c}\left(t\right)$.
One pole is at $s=0$ and two poles are at $s=\pm is_{0}$, where
$s_{0}=2\sqrt{N\left\langle g^{2}\right\rangle }$. These are given
by zeroes of the denominator of the second term in the product in
Eq.(\ref{eq:a_s}). Note that $s_{0}$ was obtained using a $1/N$-expansion
and is independent of $\xi$ in leading order. In the first subleading
$1/N$-order $s_{0}$ depends on $\xi$, \begin{equation}
s_{0}=2\sqrt{N\left\langle g^{2}\right\rangle -\frac{3}{10}g_{0}^{2}},\label{eq:s0_xig0}\end{equation}
 when $\xi=g_{0}$, and \begin{equation}
s_{0}=2\sqrt{N\left\langle g^{2}\right\rangle -\frac{1}{2}g_{0}^{2}},\label{eq:s0_xi0}\end{equation}
 when $\xi=0$. The inverse Laplace transform of the functions with
poles is a sum over the corresponding residues, $\alpha_{p}\mbox{\ensuremath{\left(t\right)}}=\sum_{s=0,\pm is_{0}}\textrm{Res}_{s}\alpha\left(s\right)e^{st}$,
and it gives\begin{equation}
\alpha_{p}\left(t\right)=\frac{1}{2}+\frac{1}{2}\cos s_{0}t+\frac{1}{N}\delta\alpha\left(\xi\right).\label{eq:ap_t}\end{equation}
 Here, the leading term is independent of $\xi$ unlike the first $1/N$-correction,
$\delta\alpha\left(g_{0}\right)=-\frac{27}{20}\left(1-\cos\left(s_{0}t\right)\right)$
and $\delta\alpha\left(0\right)=-\left(\frac{1}{4}-\frac{1}{2}\cos\left(s_{0}t\right)\right)$.
We refer to Appendix A for the calculation.

The expression in Eq. (\ref{eq:sum_even_couplings}) has four branch
points. Two are given by the square root, $s=\pm is_{1}$ where $s_{1}=\sqrt{N\left\langle g^{2}\right\rangle }$.
The remaining two are given by arctan. Solving the equation \begin{equation}
\frac{\sqrt{2}\xi\sqrt{-s^{2}-N\left\langle g^{2}\right\rangle }}{s^{2}+N\left\langle g^{2}\right\rangle -2g_{0}\left(g_{0}-\xi\right)}=\pm i\end{equation}
we find them as $s=\pm is_{2}$ where $s_{2}=\sqrt{N\left\langle g^{2}\right\rangle -\xi^{2}-2g_{0}\left(g_{0}-\xi\right)-\xi\sqrt{\xi^{2}+4g_{0}\left(g_{0}-\xi\right)}}$.
The first branch cut is chosen as a straight line between $is_{1}$
and $is_{2}$ and the second branch cut as a straight line between
$-is_{2}$ and $-is_{1}$, see Fig. \ref{fig:analytic_structure}.

The contribution to the inverse Laplace transform from the branch
cuts is a function of $\xi$. When $\xi=0$ Eq. (\ref{eq:sum_even_couplings})
has no branch points. In the $\xi\rightarrow0$ limit the arctan can be
expanded in the small parameter, then the leading term is non-zero
and contains no multivalued functions. All the higher order terms
are proportional to $\xi$ and are zero when $\xi=0$. We obtain in
this limit $\alpha_{c}\left(t\right)=0$. 

When $\xi=g_{0}$, the integral enclosing the branch cuts, \begin{equation}
\alpha_{c}\left(t\right)=\frac{32}{3N^{2}}\int_{0}^{1}dx\frac{x^{2}\cos\left(\sqrt{s_{1}^{2}-2g_{0}^{2}x^{2}}t\right)}{\left(\frac{x}{2}\ln\left(\frac{1+x}{1-x}\right)-1\right)^{2}+\left(\frac{\pi}{2}x\right)^{2}},\label{eq:ac_t}\end{equation}
contributes only to the second subleading $1/N$-order of $\alpha\left(t\right)$.
Thus, $\alpha_{c}\left(t\right)$ is beyond the accuracy of the present
calculation for all values of $\xi$ and it can be neglected compared
to the leading order correction in Eq. (\ref{eq:ap_t}).

The analytic structure of $\beta_{k}\left(s\right)$ in Eq. (\ref{eq:bs_k})
is the same as $\alpha\left(s\right)$ except that there is no pole
at $s=0$, see Fig. \ref{fig:analytic_structure}. Thus the inverse
Laplace transform also has two contributions, $\beta_{k}\left(t\right)=\beta_{k}^{p}\left(t\right)+\beta_{k}^{c}\left(t\right)$,
when $\xi>0$. One is given by the sum over just two residues, $s=\pm is_{0}$,
instead of three, $\beta_{k}^{p}\mbox{\ensuremath{\left(t\right)}}=\sum_{s=\pm is_{0}}\mathtt{Res}_{s}\beta_{k}\left(s\right)e^{st}$,
and yields\begin{equation}
\beta_{k}^{p}\left(t\right)=\frac{-ig_{k}\sin\left(s_{0}t\right)}{\sqrt{2N\left\langle g^{2}\right\rangle }}+\frac{1}{N}\delta\beta_{k}\left(\xi\right),\label{eq:bck_t}\end{equation}
where, similarly to Eq. (\ref{eq:ap_t}), only the first $1/N$ correction
depends on $\xi$ but the leading term does not, $\delta\beta_{k}\left(g_{0}\right)=-ig_{k}\left(2\left(g_{k}/g_{0}\right)^{2}-3\right)\sin\left(s_{0}t\right)/\sqrt{2N\left\langle g^{2}\right\rangle }$
and $\delta\beta_{k}\left(0\right)=i\sin\left(s_{0}t\right)/4\sqrt{2N}$,
see Appendix A for the calculation.

When $\xi=0$, the branch cuts disappear, $\beta_{k}^{c}\left(t\right)=0$,
similarly to $\alpha_{c}\left(t\right)$. When $\xi=g_{0}$ the analysis
of the branch cuts is a bit different from above
for $\alpha_{c}\left(s\right)$. There is a singularity in Eq. (\ref{eq:bs_k})
at $s=\pm i\sqrt{N\left\langle g^{2}\right\rangle -2g_{k}^{2}}$,
originating from the first term in the product in Eq. (\ref{eq:bs_k}),
which overlaps with the branch cuts. It present a difficulty if we
apply continuum approximation to the discrete form of $\beta_{k}\left(s\right)$
in the same way as we did to $\alpha\left(s\right)$. Cancellation
of this singularity by a zero in the denominator of the second term
in the product in the original discrete form, Eq. (\ref{eq:bs_k}),
simplifies the analysis. The analytic structure of $\beta_{k}\left(s\right)$
in the continuum approximation does not alter. The only difference
is a small $1/N$-correction to Eq. (\ref{eq:sum_even_couplings}).
Repeating the same steps as between Eq. (\ref{eq:sum_even_couplings})
and Eq. (\ref{eq:ac_t}) we obtain the following expression for the
integral enclosing the branch cuts,\begin{widetext}

\begin{equation}
\beta_{k}^{c}\left(t\right)=-\frac{2\sqrt{2}g_{k}^{3}g_{0}^{2}}{3N\sqrt{N\left\langle g^{2}\right\rangle }}\int_{0}^{1}dx\frac{x^{2}\left(2i\sin\left(\sqrt{N\left\langle g^{2}\right\rangle -2g_{0}^{2}x^{2}}t\right)\right)}{\left[\frac{2g_{0}^{2}x^{2}-g_{k}^{2}}{1}\frac{2}{3N}+g_{k}^{2}\left(\frac{x}{2}\ln\left(\frac{1+x}{1-x}\right)-1\right)\right]^{2}+\left[g_{k}^{2}\frac{\pi}{2}x\right]^{2}}.\end{equation}
\end{widetext}
Here, the integral can be simplified by performing an
$1/N$-expansion. This approximation is valid for the majority of
$g_{k}\gg2g_{0}/3N$ except for a small set of $g_{k}\ll2g_{0}/3N$, where
the maximum value of $\left|\beta_{k}^{c}\left(t\right)\right|\leq\frac{\sqrt{2}}{6N^{2}}$
is small as $1/\sqrt{N}$ compared to the majority of $g_{k}\gg2g_{0}/3N$.
As a result, the leading $1/N$ -term is
\begin{eqnarray}
\beta_{k}^{c}\left(t\right) & = & -\frac{\sqrt{2}}{g_{k}}\frac{4ig_{0}^{2}}{3N\sqrt{N\left\langle g^{2}\right\rangle }}I\left(t\right),\end{eqnarray}
where the dimensionless integral $ I\left(t\right)$ describes the time-decay,
\begin{equation}
I\left(t\right)=\int_{0}^{1}dx\frac{x^{2}\sin\left(\sqrt{N\left\langle g^{2}\right\rangle -2g_{0}^{2}x^{2}}t\right)}{\left(\frac{x}{2}\ln\left(\frac{1+x}{1-x}\right)-1\right)^{2}+\left(\frac{\pi}{2}x\right)^{2}},\label{eq:I_t}\end{equation}
and is independent of $k$. In contrast to $\alpha_{c}\left(t\right)$,
$\beta_{k}^{c}\left(t\right)$ does contribute to the first subleading
$1/N$-order of $\beta_{k}\left(t\right)$ and we will analyze it
below.%
\begin{figure}[pt]
\centering\includegraphics[width=0.9\columnwidth]{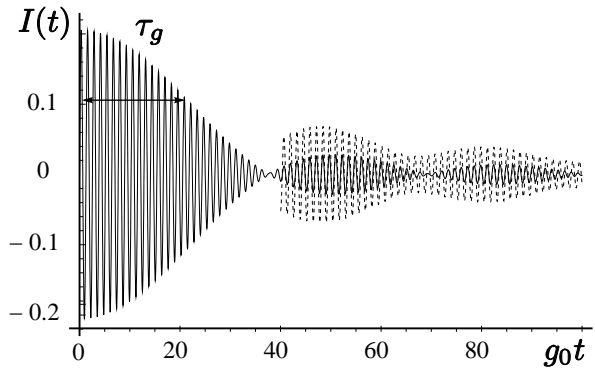}

\caption{\label{fig:I_t}Decay correction to the dynamics of the boson Eq. (\ref{eq:I_t})
for the parameter $\sqrt{N\left\langle g^{2}\right\rangle }/g_{0}=5$.
The solid line is a numerical evaluation of the integral, where $\tau_g$ marks
the time scale of the initial decay. The dashed line is the long time asymptote
Eq. (\ref{Eq: I_long_t}).}

\end{figure}

The argument of the sine in Eq. (\ref{eq:I_t}) can be expanded in $1/N$, $\sqrt{N\left\langle g^{2}\right\rangle -2g_{0}^{2}x^{2}}\approx\sqrt{N\left\langle g^{2}\right\rangle }-\frac{g_{0}^{2}x^{2}}{\sqrt{N\left\langle g^{2}\right\rangle }}$.
The leading term $\sqrt{N\left\langle g^{2}\right\rangle }t$, which
is a fast oscillating function, can be taken outside of the integral.
The second term gives a slow decay envelope. This term leads to a
significantly decay when $g_{0}^{2}t/\sqrt{N\left\langle g^{2}\right\rangle }=1$.
Thus, we estimate the decay time as \begin{equation}
\tau_{g}=\sqrt{N/\left\langle g^{2}\right\rangle },\label{tau_g}\end{equation}
see Fig. \ref{fig:I_t}. At a large time $t\gg\sqrt{N/\left\langle g^{2}\right\rangle }$,
$I\left(t\right)$ has a  power-law tail. Due to fast oscillations
of the sine the main contribution to the integral comes from $x\ll\sqrt[4]{N\left\langle g^{2}\right\rangle }/\sqrt{g_{0}^{2}t}$,
thus, the spectral function can be approximated as $x^{2}/\left(\left(\frac{x}{2}\ln\left(\frac{1+x}{1-x}\right)-1\right)^{2}+\left(\frac{\pi}{2}x\right)^{2}\right)\approx x^{2}$.
Then, the integral in Eq. (\ref{eq:I_t}) evaluates in terms of an
error function which we expand, again, in a Taylor series in powers
of $\sqrt{N\left\langle g^{2}\right\rangle }/g_{0}^{2}t\ll1$, and we
obtain\begin{multline}
I\left(t\right)=\frac{\chi}{2g_{0}t}\bigg(\cos\left(\left(\chi-\frac{1}{\chi}\right)g_{0}t\right)\\
+\frac{1}{2}\sqrt{\frac{\pi\chi}{2g_{0}t}}\left(\sin\left(\chi g_{0}t\right)+\cos\left(\chi g_{0}t\right)\right)\bigg),\label{Eq: I_long_t}\end{multline}
where $\chi=\sqrt{N\left\langle g^{2}\right\rangle }/g_{0}$. The shape
of the asymptote is in qualitative agreement with the explicit numerical
evaluation of Eq. (\ref{eq:I_t}), however, the overall amplitude is
different by a factor of 3, see Fig. \ref{fig:I_t}, as we neglected
the logarithmic singularity at $x=1$ in the initial integral.

In the sum of two amplitudes $\alpha\left(t\right)$ and $\beta_{k}\left(t\right)$
the decay shows up only in the fist subleading $1/N$-order
when the number of spins is large. The particular form of the decay function
is rather involved and is not displayed here.

The decay on the short time scale $t<\tau_{g}$ is essentially non-exponential,
see Fig. \ref{fig:I_t}. The particular shape depends on the particular
set of the strongest coupling constants $g_{j}$. However, an estimate
of the time scale $\tau_{g}\sim\sqrt{N/\left\langle g^{2}\right\rangle }$
is independent of $Q\left(g\right)$, as it is based on an $1/N$-expansion
only, i. e. the distance between the branch points in Fig. \ref{fig:I_t}
is smaller by $1/N$ compared to the distances between the branch points
and the poles. The power law tail exists due to a bound on the smallest
$g_{j}$. The power and the numerical prefactor in Eq. (\ref{Eq: I_long_t})
depend on a particular $Q\left(g\right)$, especially on the distribution
of the smallest $g_{j}$'s as they are responsible for the long time
behavior.

The time-dependent occupation number of the boson can also be expanded
into a $1/N$-series. The leading term depends on $\xi$ only through
the effective coupling $\sqrt{\left\langle g^{2}\right\rangle }$,

\begin{equation}
\left\langle n(t)\right\rangle =2\cos^{2}\left(\frac{s_{0}t}{2}\right)+\frac{1}{N}\delta n\left(t\right).\label{eq:boson_number_expansion}\end{equation}
The leading $1/N$-correction, $\delta n\left(t\right)=4\alpha\left(t\right)\delta\alpha\left(t\right)+2\sum_{j}\beta_{j}\left(t\right)\delta\beta_{j}\left(t\right)$,
is qualitatively different for $\xi=0$ and $\xi>0$. When the coupling
constants are homogeneous, $\xi=0$, \begin{equation}
\delta n\left(t\right)=-\left(2\left(1+\cos\left(s_{0}t\right)\right)+\frac{1}{8}\left(1-\cos\left(2s_{0}t\right)\right)\right)\end{equation}
is an oscillatory function where the second harmonic with the doubled
frequency $2s_{0}$ appears in addition to the main frequency $s_{0}$
of the leading term. For the case of maximally inhomogeneous coupling constants, $\xi=g_{0}$,
the function \begin{equation}
\delta n\left(t\right)=1.8\sin^{2}\left(s_{0}t\right)-2\sin\left(s_{0}t\right)I\left(t\right) \label{delta_n_g0}\end{equation}
contains a decaying contribution, where $I\left(t\right)$ is the decay function
from Eq. (\ref{eq:I_t}). Up to the time scale $\tau_{g}$ the term
$\sin\left(s_{0}t\right)I\left(t\right)\sim\sin\left(s_{0}t\right)\cos\left(s_{1}t\right)$
is a harmonic mode with the third frequency $s_{0}+s_{1}$ in addition
to  $s_{0}$ and $2s_{0}$. This mode can only be observed if
the short-time regime with $t<\tau_{g}$ is accessible to measurement.

\subsection{Sawtooth distribution function $Q_{2}\left(g\right)$ }

Here we study another distribution of the coupling constants. Assuming
that there are more spins at the nodes of the cavity mode, so that the stronger
coupling constants are more favorable, we consider a sawtooth-like distribution
function with its maximum at the largest coupling strength $g_{0}$, $Q_{2}\left(g\right)=2g/g_{0}^{2}\theta\left(-g+g_{0}\right)\theta\left(g\right)$,
see Fig. \ref{fig:P_gs}b. 

Replacing the sums in Eqs. (\ref{eq:bs_k}, \ref{eq:a_s}) by
integrals, $\sum_{j}\dots\rightarrow N\int_{0}^{\infty}dgQ\left(g\right)\dots$,
and using $Q_{2}\left(g\right)$ we get\begin{widetext} \begin{equation}
\left\langle \frac{g_{j}^{2}N}{-s^{2}+2g_{j}^{2}-N\left\langle g^{2}\right\rangle }\right\rangle =\frac{N}{2}\left(1-\frac{s^{2}+N\left\langle g^{2}\right\rangle }{2g_{0}^{2}}\ln\left(\frac{s^{2}+N\left\langle g^{2}\right\rangle }{s^{2}+N\left\langle g^{2}\right\rangle -2g_{0}^{2}}\right)\right),\label{eq:sum_tooth_couplings}\end{equation}
 \end{widetext}where $\left\langle g^{2}\right\rangle =g_{0}^{2}/2$.

The analytic structure of Eqs. (\ref{eq:bs_k}, \ref{eq:a_s}) with
the sum from the above equation is the same as with the sum from Eq.
(\ref{eq:sum_even_couplings}) obtained using $Q_{1}\left(g\right)$.
There are three poles ($\alpha\left(s\right)$ has three poles and
$\beta_{k}\left(s\right)$ has only two as in the previous subsection)
and two branch cuts, see Fig. \ref{fig:analytic_structure}. Thus,
the inverse Laplace transforms of $\alpha\left(s\right)$ and $\beta_{k}\left(s\right)$
also have two contributions, i.e. $\alpha\left(t\right)=\alpha_{p}\left(t\right)+\alpha_{c}\left(t\right)$
and $\beta_{k}\left(t\right)=\beta_{k}^{p}\left(t\right)+\beta_{k}^{c}\left(t\right)$.
Similarly to the previous subsection, there is a pole at $s=0$ and
there are two poles at $s=\pm is_{0}$, where $s_{0}=2\sqrt{N\left\langle g^{2}\right\rangle }$
agrees in  leading $1/N$-order  with what was
obtained in the previous subsection. The four branch points, which
are due to the logarithm in Eq. (\ref{eq:sum_tooth_couplings}), are
found from \begin{equation}
\frac{s^{2}+N\left\langle g^{2}\right\rangle }{s^{2}+N\left\langle g^{2}\right\rangle -2g_{0}^{2}}=0,\infty,\end{equation}
as $s=\pm is_{1,2}$ where $s_{1}=\sqrt{N\left\langle g^{2}\right\rangle }$
and $s_{2}=\sqrt{N\left\langle g^{2}\right\rangle -2g_{0}^{2}}$.
These also agree with what we have already found in the previous
subsection when the coupling constants were maximally inhomogeneous, i.e. $\xi=g_{0}$.

The sums over the residues give the main contribution to the inverse
Laplace transforms. The leading $1/N$-terms in $\alpha\left(t\right)$
and $\beta_{k}\left(t\right)$ agree with the leading terms in
Eq. (\ref{eq:ap_t}) and Eq. (\ref{eq:bck_t}), where $\left\langle g^{2}\right\rangle $
has to be calculated using the sawtooth distribution function $Q_{2}\left(g\right)$
instead of $Q_{1}\left(g\right)$. The constributions from the branch
cuts also appear only in the first subleading $1/N$ order.
The main features of the time decay are similar to that of Eq. (\ref{eq:I_t}).
Indeed, the decay time $\tau_{g}$ and the frequency of
the fast oscillating term in Eq. (\ref{eq:I_t}) for times
$t<\tau_{g}$ result from the same branch points, $\pm is_{1,2}$,
as in the previous subsection.

As the amplitudes $\alpha\left(t\right)$ and $\beta_{k}\left(t\right)$
are similar to the ones in the previous subsection,  the boson  number 
$\left\langle n(t)\right\rangle$
is also given by Eq. (\ref{eq:boson_number_expansion}). The leading
$1/N$-term depends on $Q_{2}\left(g\right)$ only through the effective
coupling constant $\sqrt{\left\langle g^{2}\right\rangle }$, and the
leading $1/N$-correction contains a decay term.

\section{Subspace of two and three excitations}

In this section we compare numerically the solution of the 
Schr\"odinger equation with the one of the classical equations
of motion Eqs. (\ref{eq:C_j}, \ref{eq:a}) for $p=2,3$ excitations.
We start from writing down the Schr\"odinger equation in these two
subspaces explicitly.

The time evolution of two excitations is restricted to the subspace
with $c=-N/2+2$ and is described by the general state Eq. (\ref{eq:wave_function}).
The Schr\"odinger equation for an arbitrary set of $\epsilon_{j}$
and $g_{j}$ in this subspace is similar to Eq. (\ref{eq:dynamic_equations_t}),
\begin{eqnarray}
-i\dot{\alpha} & = & \sqrt{2}\sum_{j}g_{j}\beta_{j}\label{eq:Schrodinger_EoM_p2}\\
-i\dot{\beta}_{k} & = & \left(\epsilon_{k}-\omega\right)\beta_{k}+\sqrt{2}g_{k}\alpha+\sum_{j<k}g_{j}\gamma_{kj}+\sum_{j>k}g_{j}\gamma_{jk}\nonumber \\
-i\dot{\gamma}_{kl} & = & \left[\left(\epsilon_{k}-\omega\right)+\left(\epsilon_{l}-\omega\right)\right]\gamma_{kl}+\left(g_{k}\beta_{l}+g_{l}\beta_{k}\right)\left(1-\delta_{kl}\right),\nonumber \end{eqnarray}
with the initial condition $\alpha\left(0\right)=1$, $\beta_{k}\left(0\right)=0$,
and $\gamma_{kl}\left(0\right)=0$.

The subspace of three excitations is labeled by $c=-N/2+3$ and is
described by the general state,\begin{widetext}

\begin{equation}
\left|\Psi\left(t\right)\right\rangle
=\alpha\left(t\right)\left|3,\Downarrow\right\rangle +\sum_{j}\beta_{j}\left|2,\Downarrow\uparrow_{j}\right\rangle +\sum_{i>j}\gamma_{ij}\left|1,\Downarrow\uparrow_{i}\uparrow_{j}\right\rangle +\sum_{i>j>r}\eta_{rij}\left|0,\Downarrow\uparrow_{r}\uparrow_{i}\uparrow_{j}\right\rangle ,\end{equation}
\end{widetext}
where $\alpha\left(t\right)$, $\beta_{j}\left(t\right)$,
$\gamma_{ij}\left(t\right)$, and $\eta_{rij}\left(t\right)$ are
the normalized amplitudes, $\left|\alpha\left(t\right)\right|^{2}+\sum_{j}\left|\beta_{j}\left(t\right)\right|^{2}+\sum_{i>j}\left|\gamma_{ij}\right|^{2}+\sum_{i>j>r}\left|\eta_{rij}\right|^{2}=1$,
of the state with three bosonic excitations, a state with two bosonic
excitations and the $j^{th}$ spin excited, a state with one bosonic
excitation and the $i^{th}$ and $j^{th}$ spins excited, and a state
with no bosonic excitations and the $i^{th}$, $j^{th}$ and $r^{th}$
spins excited. The amplitude $\gamma_{ij}$ is defined such that $\gamma_{ij}=0$
if $j\geq i$, and $\eta_{rij}$ is defined such that $\eta_{rij}=0$
if the inequality $i>j>r$ is not satisfied. The Schr\"odinger equation
in this subspace is \begin{widetext} \begin{eqnarray}
-i\dot{\alpha} & = & \sqrt{3}\sum_{j}g_{j}\beta_{j},\label{eq:Schrodinger_EoM_p3}\\
-i\dot{\beta}_{k} & = & \left(\epsilon_{k}-\omega\right)\beta_{k}+\sqrt{3}g_{k}\alpha+\sqrt{2}\sum_{j<k}g_{j}\gamma_{kj}+\sqrt{2}\sum_{j>k}g_{j}\gamma_{jk},\nonumber \\
-i\dot{\gamma}_{kl} & = & \left[\left(\epsilon_{k}-\omega\right)+\left(\epsilon_{l}-\omega\right)\right]\gamma_{kl}+\sqrt{2}\left(g_{k}\beta_{l}+g_{l}\beta_{k}\right)\left(1-\delta_{kl}\right)+\sum_{j>k>l}g_{j}\eta_{jkl}+\sum_{k>j>l}g_{j}\eta_{kjl}+\sum_{k>l>j}g_{j}\eta_{klj},\nonumber \\
-i\dot{\eta}_{klm} & = & \left[\left(\epsilon_{k}-\omega\right)+\left(\epsilon_{l}-\omega\right)+\left(\epsilon_{m}-\omega\right)\right]\gamma_{kl}+g_{k}\gamma_{lm}+g_{l}\gamma_{km}+g_{m}\gamma_{lm}\nonumber \end{eqnarray}
\end{widetext} 
with the initial conditions $\alpha\left(0\right)=1$,
$\beta_{k}\left(0\right)=0$, and $\gamma_{kl}\left(0\right)=0$. The
physical observable of  interest is again the time-dependent boson number
 which can be expressed in terms of the amplitudes
$\alpha\left(t\right)$, $\beta_{j}\left(t\right)$, and $\gamma_{ij}\left(t\right)$
as \begin{equation}
\left\langle n(t)\right\rangle =3\left|\alpha\left(t\right)\right|^{2}+2\sum_{j}\left|\beta_{j}\left(t\right)\right|^{2}+\sum_{j}\left|\gamma_{ij}\left(t\right)\right|^{2}.\label{eq:n_abg}\end{equation}
\begin{figure*}
\centering\includegraphics[width=0.85\columnwidth]{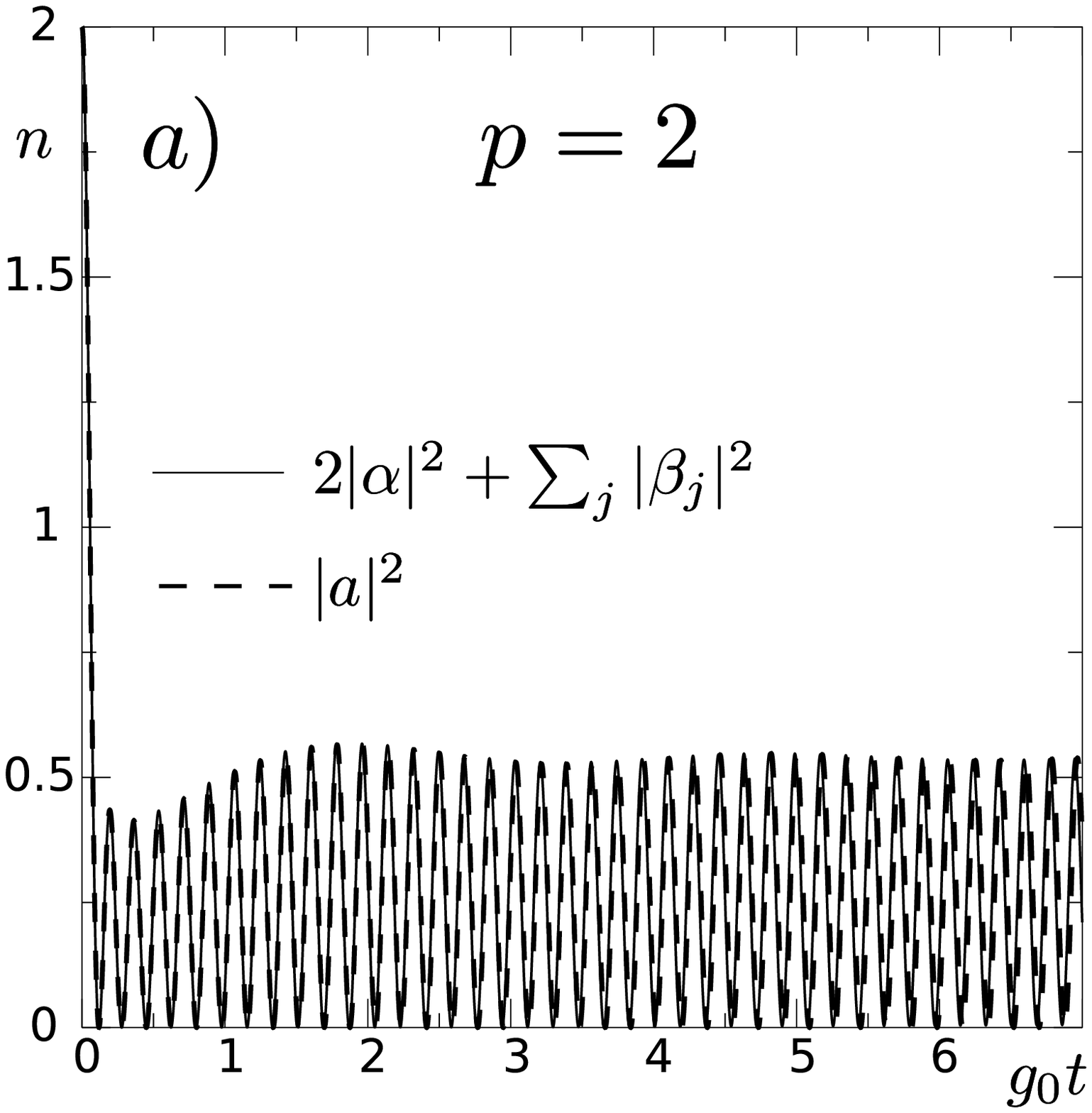}\qquad{}\includegraphics[width=0.85\columnwidth]{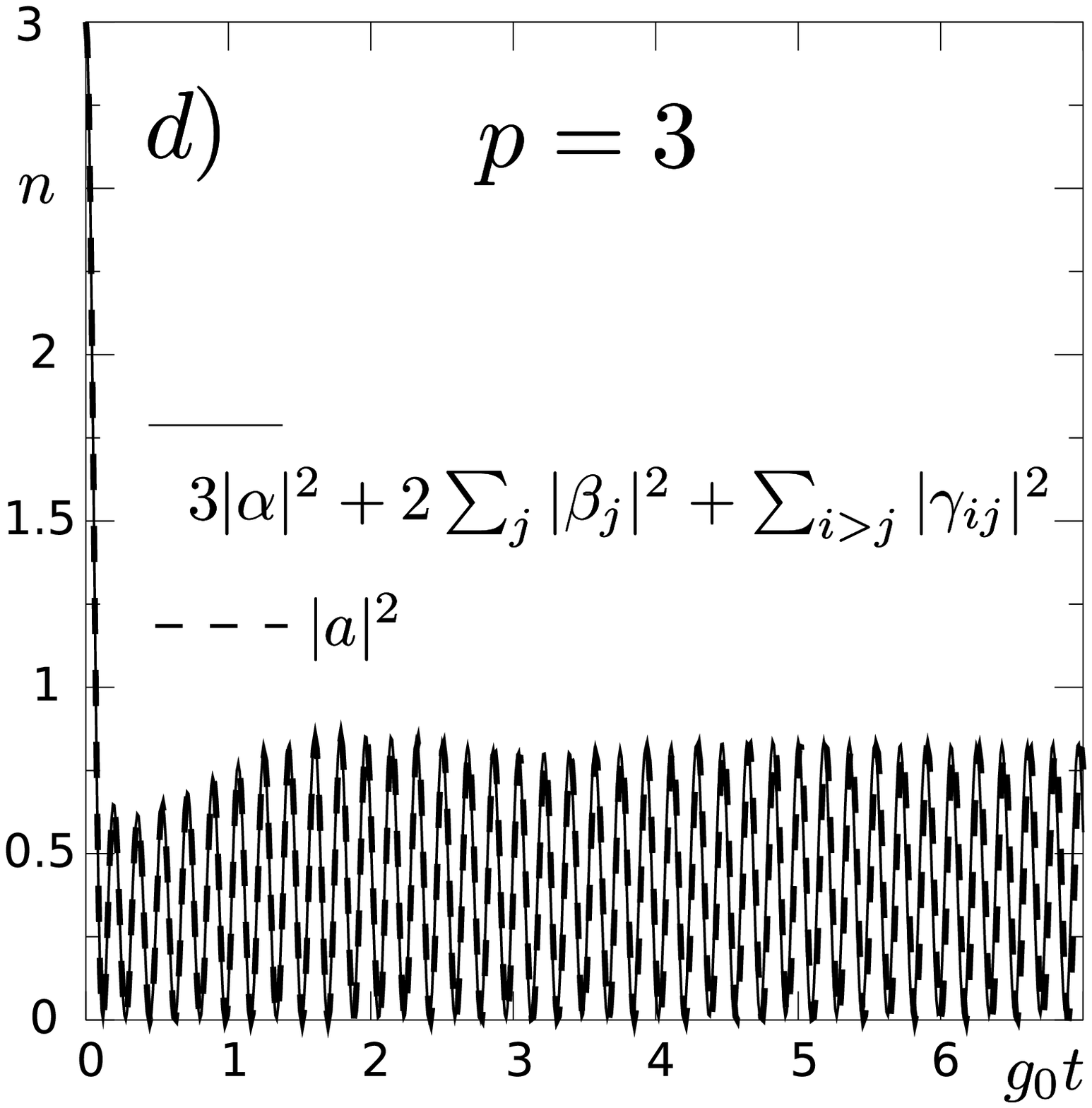}\medskip{}

\includegraphics[width=0.85\columnwidth]{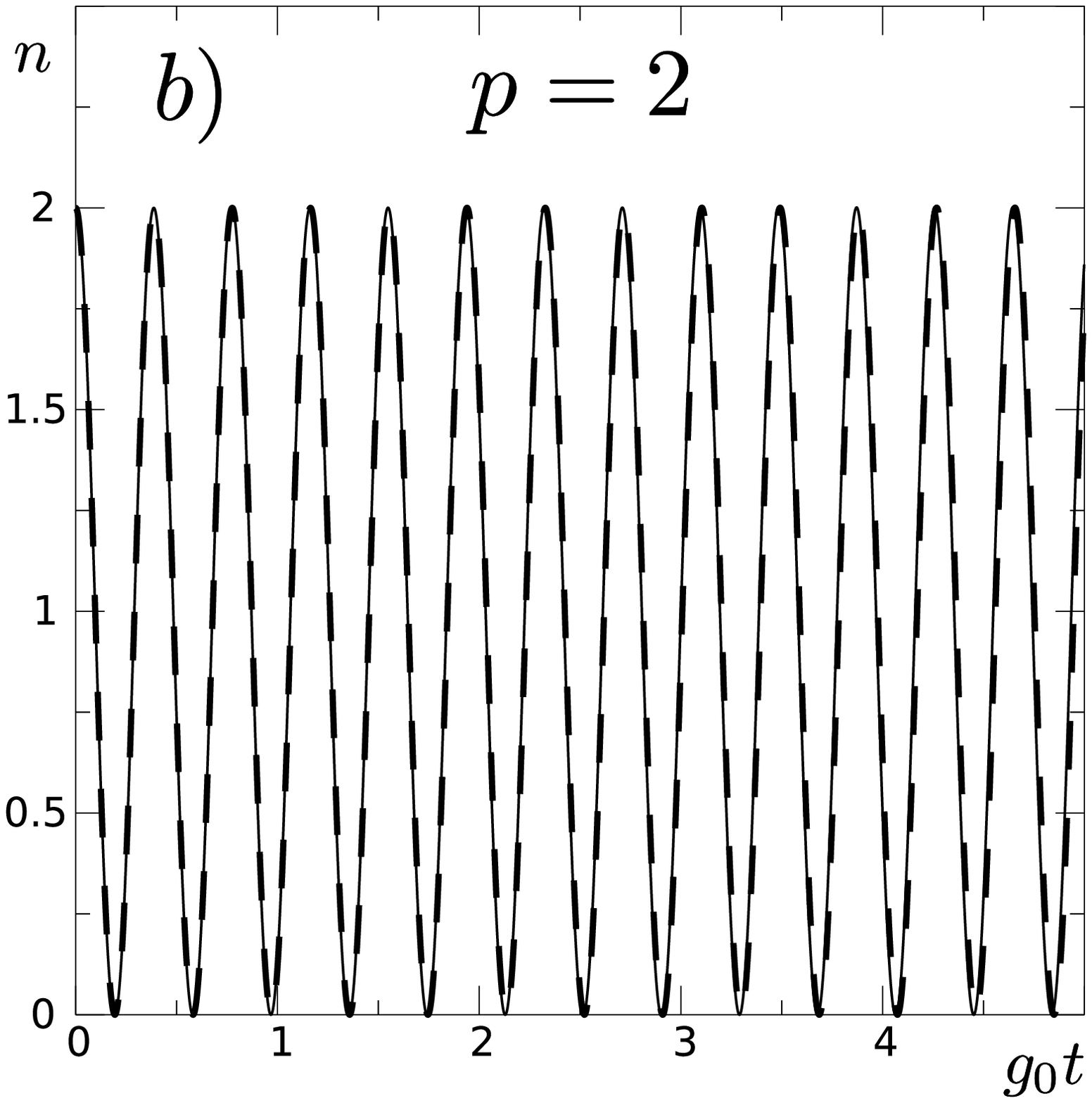}\qquad{}\includegraphics[width=0.85\columnwidth]{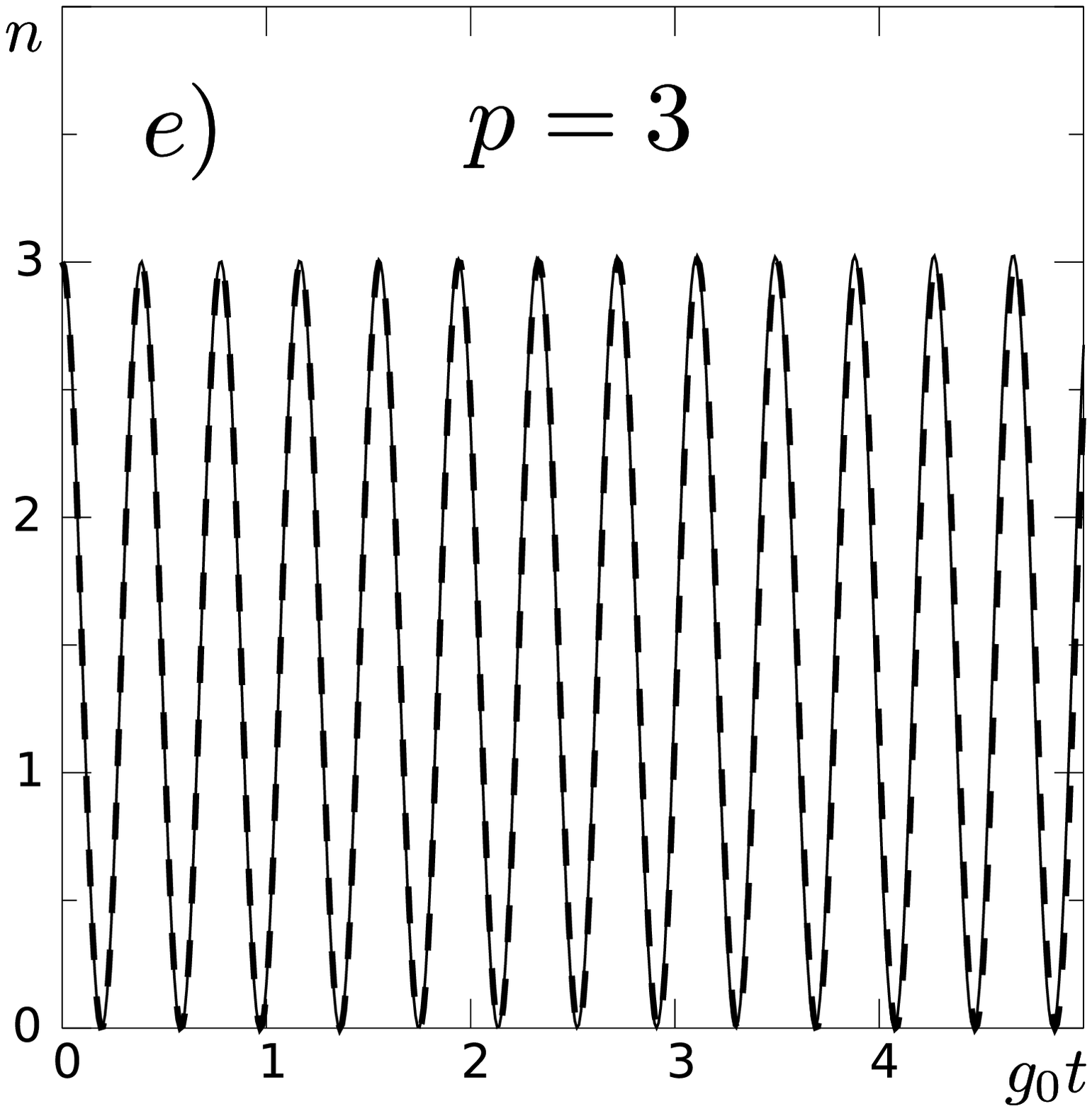}\medskip{}
\includegraphics[width=0.85\columnwidth]{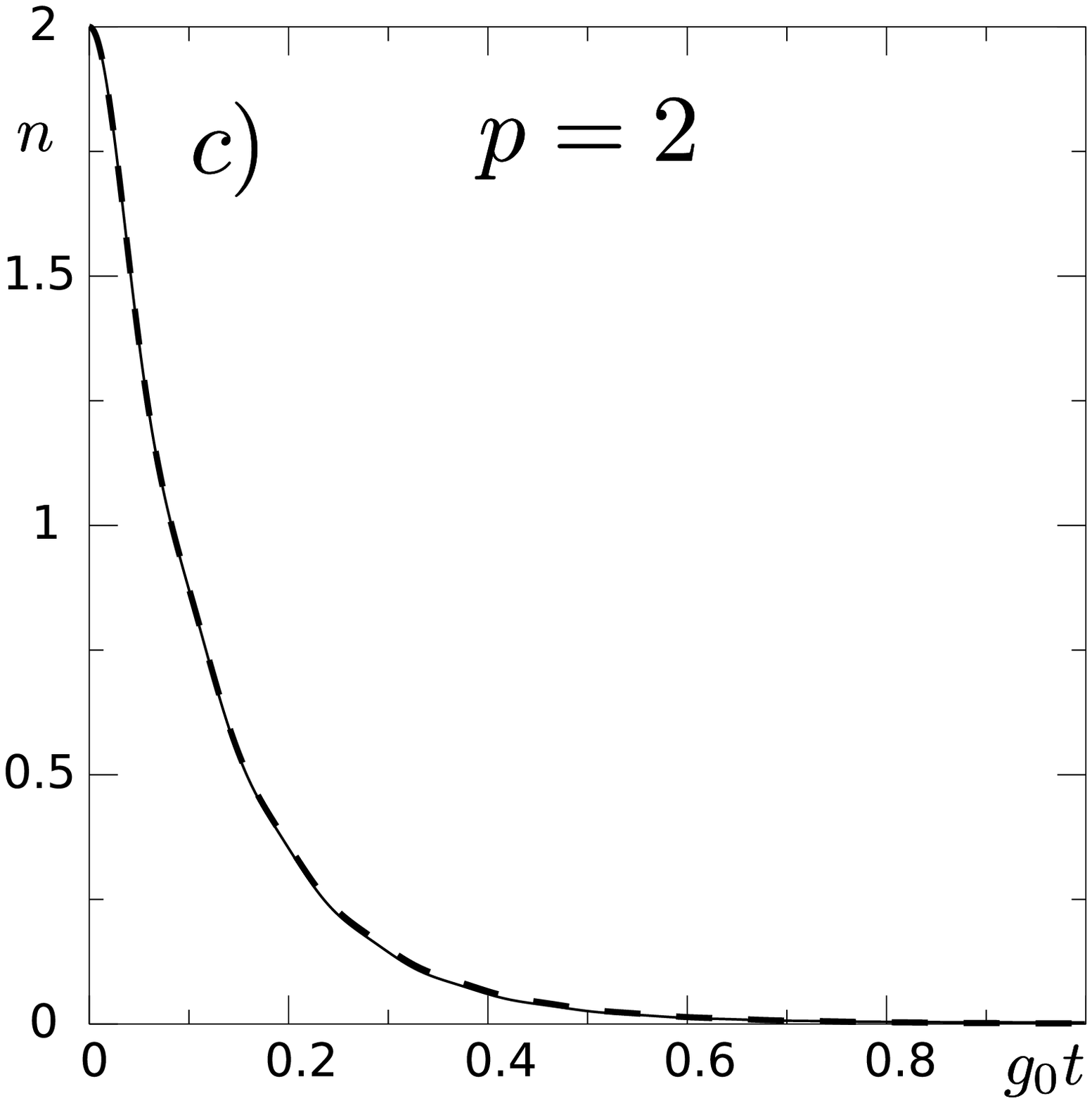}\qquad{}\includegraphics[width=0.85\columnwidth]{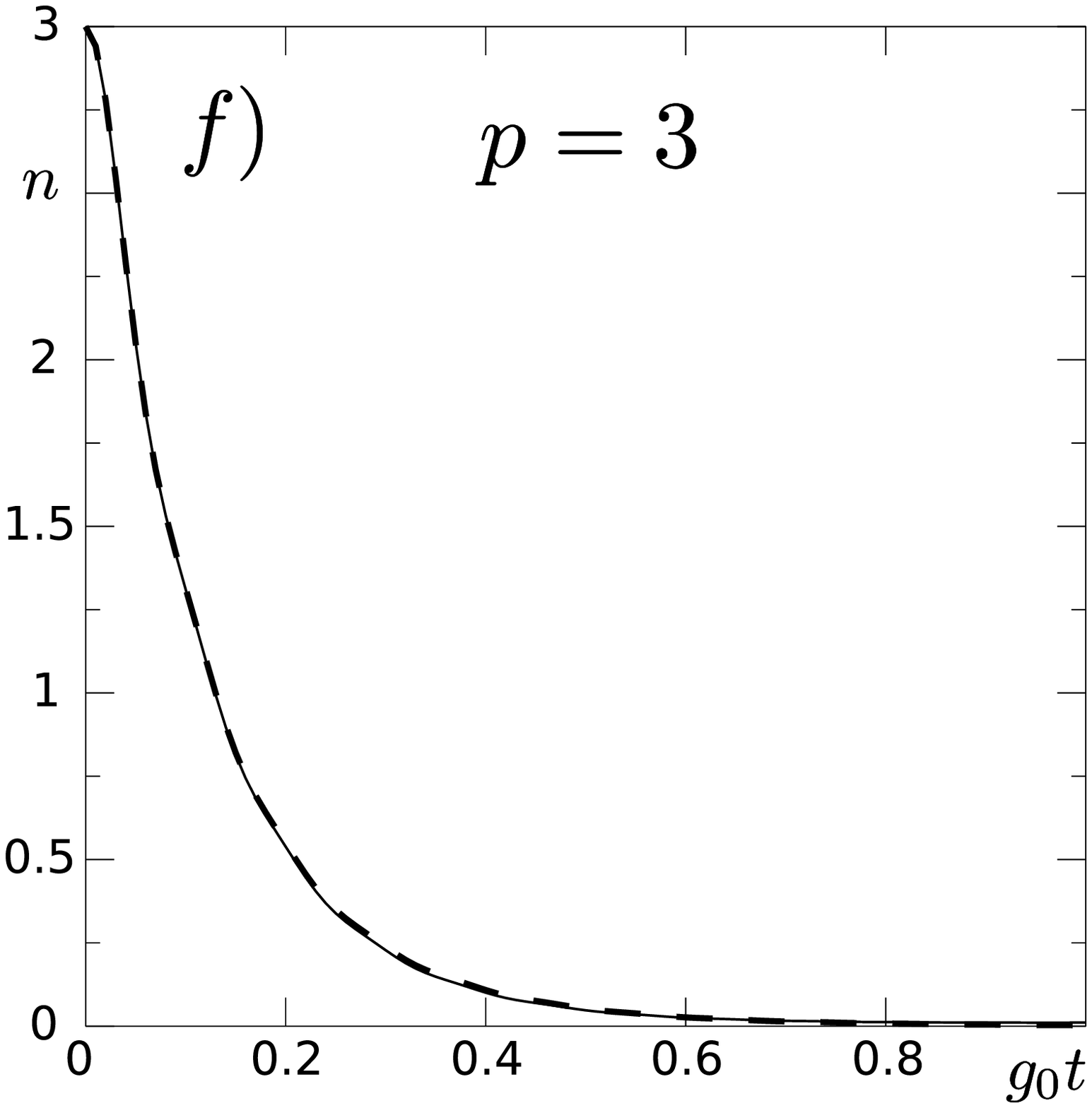}

\caption{\label{fig:classical_quantum_numerics}Numerical solutions of the Schr\"odinger
Eqs. (\ref{eq:Schrodinger_EoM_p2}, \ref{eq:Schrodinger_EoM_p3})
and classical equations of motion Eqs. (\ref{eq:C_j}, \ref{eq:a})
for a large number of spins, $N=200$, and $p=2,3$ number of excitations.
The inhomogeneities are characterized by the distributions $P\left(\epsilon\right)$
and $Q_{1}\left(g\right)$. In plots a) and d): $\Delta/\Omega=2.2$
and $\xi=0$; in plots b) and e): $\Delta=0$ and $\xi=g_{0}$; in
plots c) and f): $\Delta/\Omega=10$ and $\xi=0$.}

\end{figure*}

Unlike before for $p=1$, the classical equations of motion
Eqs. (\ref{eq:C_j}, \ref{eq:a}) and the Schr\"odinger Eqs. (\ref{eq:Schrodinger_EoM_p2},
\ref{eq:Schrodinger_EoM_p3}) are not equivalent in the subspaces
of $p=2,3$. Solving these equations numerically, we compare the time-dependent
boson number, given by Eqs. (\ref{eq:n_ab}, \ref{eq:n_abg}),
for $p=2,3$ with the  square modulus of the classical
field $a\left(t\right)$ obtained from  Eqs. (\ref{eq:C_j},
\ref{eq:a}). In the large-$N$ limit the solutions of both classical
and quantum equations have exactly the same form.

In Fig. \ref{fig:classical_quantum_numerics} we give a detailed comparison
of the different types of inhomogeneities characterized by  the distribution
functions $P\left(\epsilon\right)$ and $Q_{1}\left(g\right)$ for
a large number of spins, i.e. $N=200$. The solutions of Eq. (\ref{eq:Schrodinger_EoM_p2})
and Eqs. (\ref{eq:C_j}, \ref{eq:a}) with $p=2$ are compared in
Fig. \ref{fig:classical_quantum_numerics} a)-c), and the solutions
of Eq. (\ref{eq:Schrodinger_EoM_p3}) and Eqs. (\ref{eq:C_j}, \ref{eq:a})
with $p=3$ are compared in Fig. \ref{fig:classical_quantum_numerics}
d)-f). The regime with inhomogeneous coupling constants only, characterized by $Q_{1}\left(g\right)$
from Eq. (\ref{eq:Q1_g}) with $\xi=g_{0}$,
and $P\left(\epsilon\right)$
from Eq. (\ref{eq:P_eps}) with $\Delta=0$, is shown in Fig.
\ref{fig:classical_quantum_numerics} b) and e). The regime of strongly
inhomogeneous Zeeman energies, $\Delta/\Omega=10$, is shown
in Fig. \ref{fig:classical_quantum_numerics} c) and f). Finally, an intermediate regime with $\Delta/\Omega=2.2$
 is shown in Fig. \ref{fig:classical_quantum_numerics} a) and d). 

As the classical and quantum solutions coincide in all regimes, the classical equations  can be used to find the  time-dynamics of the boson occupation number. This is quite remarkable since  the classical equations are significantly simpler to solve than the  Schr\"odinger equation, both analytically and numerically.
When the number of excitation is
small, $p\ll N$, the classical equations  can be mapped to
the Schr\"odinger equation in the one-excitation subspace $p=1$
in  leading $1/N$-order, see Eqs. (\ref{eq:a}, \ref{eq:Cm_j}).
The Schr\"odinger equation for this case was already
solved. A larger number p of excitations changes only the initial condition
of Eq. (\ref{eq:eqs_of_motion_1}) from $\alpha\left(0\right)=1$
to $\alpha\left(0\right)=\sqrt{p}$. This difference can be accounted for by a simple rescaling by $p$ of the bosonic occupation number
$\left\langle n \left(t\right)\right\rangle $ that was obtained in the 
single-excitation subspace for all regimes. The explicit
results were given in Sec. III, see Eqs.
(\ref{eq:1_excitation_1_frequency}, \ref{eq:1_excitation_exp}),
but we do not show them in Fig. \ref{fig:classical_quantum_numerics}. 

In conclusion, the analysis in Ref. \onlinecite{OT_Loss} is also
applicable to the case with more than one excitation, provided $p\ll N$ and $N\gg 1$.

\section{Applicability of the classical approximation}

The classical approximation in the few-excitation sector is exact
when the number of spins $N$ is infinite. For  finite but still
large N's, i.e. $N\gg1$, the time-evolution of the classical system deviates from 
the quantum system by a small amount. The goal of this section
is to analyze these finite-size deviations  \cite{diffusive_systems}. 

\subsection{Ehrenfest time}
One way to quantify the difference between the classical 
and the quantum solution is to identify the Ehrenfest time
 $\tau_E$ at which they deviate significantly from each other. 
To this end, we compare the boson number
$\left\langle n(t)\right\rangle$,
obtained from the classical equations (in the few-excitation approximation), Eqs. (\ref{eq:a}, \ref{eq:Cm_j}), with
the exact quantum solution in the two-excitation subspace
obtained in Sec. V. The solution to the classical equations 
Eq. (\ref{eq:1_excitation_1_frequency}) in this regime is $n=2\cos^{2}\left(\Omega t\right)$,
where $\Omega=\sqrt{N\left\langle g^{2}\right\rangle }$. From Eq.
(\ref{eq:boson_number_expansion}) the solution to the Schr\"odinger equation
is $\left\langle n\right\rangle =2\cos^{2}\left(s_{0}t/2\right)$
in  leading $1/N$-order. Both solutions are  single harmonic modes
with frequencies that also match in leading $1/N$-order, where $s_{0}=2\sqrt{N\left\langle g^{2}\right\rangle }$.
In first subleading $1/N$-order, the correction to $s_{0}$ depends
explicitly on $Q_{1}\left(g\right)$, see Eqs. (\ref{eq:s0_xig0},
\ref{eq:s0_xi0}). For $\xi=0$ the expansion of Eq. (\ref{eq:s0_xi0})
gives $s_{0}=2\sqrt{N\left\langle g^{2}\right\rangle }-\sqrt{4\left\langle g^{2}\right\rangle /N}$,
and for $\xi=g_{0}$ expanding Eq. (\ref{eq:s0_xig0}) we get $s_{0}=2\sqrt{N\left\langle g^{2}\right\rangle }-3\sqrt{3\left\langle g^{2}\right\rangle /10N}$.
Thus, at the time scale
\begin{equation}
\tau_{E}=\sqrt{\frac{N}{\left\langle g^{2}\right\rangle }} \end{equation}
the phase difference between the two harmonic modes, with frequencies
$\Omega$ and $s_{0}/2$, is comparable to $2\pi$. Hence,  the difference between
classical and quantum solutions   is significant for any value of $\xi$ at  this time
scale $\tau_{E}$, to which we refer as Ehrenfest time.
The numerical prefactor depends
on the particular type of inhomogeneity. The explicit calculations in Sec. V
for typical  distribution functions  show that these numerical prefactors
are of order one.

There is also a $1/N$-correction to the amplitude of $\langle n(t)\rangle$ coming from the Schr\"odinger equation, see Eq. (\ref{delta_n_g0}). This correction contains a decaying contribution (proportional to $I(t)$ given in Eq. (\ref{eq:I_t})) which,
again, marks a qualitative difference between the quantum and  classical time-dynamics in the two-excitation subspace. 
In particular, $I(t)$ decays at the characteristic time scale $\tau_g$ given in Eq. (\ref{tau_g}) which is equal to the Ehrenfest time $\tau_E$ introduced above. Thus, we see that this difference in the amplitude (although it is only a 1/N-correction) is another manifestation of the quantum nature of the system where the time-dynamics for times exceeding the Ehrenfest time  can be described correctly only by the Schr\"odinger equation (and not by the classical one).

So far we have been using the approximate classical Eqs. (\ref{eq:a}, \ref{eq:Cm_j})
 for few excitations, thereby neglecting the deviations of the z-component
of the classical spins from $C_{j}^{z}=-1/2$. To estimate the quality
of this approximation, we use the result obtained in Ref. \onlinecite{JaynesCummings}
for the homogeneous classical system. The solution of the 
unapproximated Eqs. (\ref{eq:C_j}, \ref{eq:a})
with $g_{j}\equiv g_{0}$ and $\epsilon_{j}\equiv\omega$ is an elliptic
function of time \cite{JaynesCummings}. When the number of excitations is small, $p\ll N$, this elliptic function
can be expanded into a harmonic series with a leading term that
reproduces the solution of the  approximate
Eqs. (\ref{eq:a}, \ref{eq:Cm_j}). The frequency of the leading harmonic
term matches the frequency $\Omega=\sqrt{N}g_{0}$ from Eq. (\ref{eq:1_excitation_1_frequency})
in leading $1/N$-order, but it also contains corrections
on the order of $g_{0}/\sqrt{N}$ like Eq. (\ref{eq:boson_number_expansion}).
Such corrections, however, are irrelevant as they become only sizable at  
the Ehrenfest time $\tau_{E}$--the time beyond which the classical solution fails and the true time-dynamics
must be described anyway by the Schr\"odinger equation. 

\subsection{ Initial spin excitations}
Up to now we have  focussed on a particular initial condition with
 excitations being  initially present only in the boson mode. In contrast,
  a different
initial condition was considered in Ref. \onlinecite{Cummings},
whereby the dynamics starts from an initial 
state with no boson present but, say, with the $i^{th}$ spin excited.  Considering
homogeneous systems, it was found that 
during the time-evolution
this $i^{th}$ spin remains excited if the total number of spins is
large, regardless of how strong the spin-boson coupling is. The corresponding
expectation value is $\left\langle 0,\Downarrow\uparrow_{i}\middle|S_{i}^{z}
\left(t\right)\middle|0,\Downarrow\uparrow_{i}\right\rangle =
1/2-\left(1-\cos\left(\Omega t\right)\right)/N$.
This result was associated with the effect of {}``radiation trapping'',
and it can also be obtained with the classical approximation. The
corresponding initial condition $\mathbf{C}_{j\neq i}\left(0\right)=\left(0,0,-1/2\right)$,
$\mathbf{C}_{i}\left(0\right)=\left(0,0,1/2\right)$, and $a\left(0\right)=0$
is a fixed point of the classical Eqs. (\ref{eq:C_j}, \ref{eq:a}).
Indeed, the effective magnetic field for each spin $\mathbf{B}_{j}=\left(0,0,\epsilon_{j}-\omega\right)$
has only a z-component, and therefore the vector-product of two parallel
vectors vanishes, i.e. $\mathbf{B}_{j}\times\mathbf{C}_{j}=0$. The dynamics
of the classical field $a$ is frozen as $C_{j}^{-}=0$. Quantum 
corrections to this result show up only in the first $1/N$-correction.
 Thus, the classical approximation is also
valid for a different initial condition in the few-excitation regime.

\section{Conclusions}

In this paper we have shown that the solution to the classical Hamilton
equations of motion of the inhomogeneous Dicke model coincide with
the solution of the time-dependent Schr\"odinger equation when the
number of spins $N$ is large and the number of excitations $p$ is
small, $p\ll N$. For a single excitation the leading $1/N$-order 
of the classical solution coincides with the quantum solution.
For a few excitations such correspondence does not hold but for $p=2,3$
excitations the numerical solutions of both classical equations of
motion and Schr\"odinger equation coincide when the number of spins
is large. It is plausible to conjecture that the same correspondence 
holds for $p>3$  in leading $1/N$-order.

To assess the validity of the classical approximation for $p>1$ excitations,
we have solved the Schr\"odinger equation exactly in the two-excitation
subspace with inhomogeneous coupling constants only and compared the result
with the classical solutions. For  large $N$, we performed
an $1/N$-expansion of the solution to the Schr\"odinger equation and
recovered the classical solution in leading order. Subleading $1/N$-corrections
cause small deviations of the classical from the quantum solution,
that, at a large time scale, make the difference between the two significant.
This defines the Ehrenfest time that we identify in the limit of $p\ll N$
as $\tau_{E}=\sqrt{N/\left\langle g^{2}\right\rangle }$. 

Analyzing the solution to the Schr\"odinger equation for $p=2$, we compared
it with the solutions to the Schr\"odinger equation for $p=1$. We have
found that the boson occupation number in the two-excitation subspace
exhibits a multi-frequency dynamics due to the inhomogeneous couplings
only, which, unlike in the single-excitation subspace, can lead to
a decay in the limit of large $N$. But the leading term of an $1/N$-expansion 
recovers the single-frequency dynamics. The decay due to
the inhomogeneity shows up only in the first subleading $1/N$-correction.
We find that this contribution is an oscillatory mode with frequency
$\frac{3}{2}\Omega$ and a slow decay envelope. The decay  is
essentially non-exponential with a long power-law tail, and the decay
time is $\tau_{g}\sim\sqrt{N/\left\langle g^{2}\right\rangle }$, where
$\sqrt{\left\langle g^{2}\right\rangle }$ is a characteristic coupling,
the numerical prefactor is of order one for the special case of  uniformly distributed
coupling constants. 

The decay due to an inhomogeneous coupling to a spin bath, which is
unavoidable as the spins are located at different positions of
the cavity mode (with different amplitudes of the electromagnetic field),
is similar to the decay of an electron spin coupled to a bath of nuclear
spins through the   hyperfine interaction \cite{KhaetskiiLossGlazman,CoishLoss}.
In the dynamics of a cavity mode this mechanism can be neglected when
only a few excitations are present in the system (for instance, in the
few-photon spectroscopy experiments in Ref. \onlinecite{Esslinger}), but may lead
to a significant decay in a system with many excitations
present   initially (such as, for instance, in a bath of nuclear
spins coupled to a cavity).

\section{Acknowledgments}
We thank  J. von Delft and A. Imamo\=glu for discussions. 
We acknowledge
support from the Swiss NSF, NCCR Nanoscience Basel, JST
ICORP, and DARPA QuIST.

\appendix

\section{$1/N$ corrections to $\alpha_{p}\left(t\right)$ and $\beta_{k}^{p}\left(t\right)$}

In this appendix we calculate the first $1/N$ correction to the pole
contributions to the inverse Laplace transform of $\alpha\left(t\right)$
and $\beta_{k}\left(t\right)$ from Sec. V. 

If $\xi=0$, i.e. for homogeneous coupling constants, the solution
to  Eq. (\ref{eq:dynamic_equations_s}) simplifies. Substituting
$g_{j}=g_{0}$ into Eqs. (\ref{eq:bs_k}, \ref{eq:a_s}), we get
\begin{equation}
\beta_{k}\left(s\right)=\frac{i\sqrt{2}g_{0}}{s^{2}+\left(4N-2\right)g_{0}^{2}},\end{equation}
 and \begin{equation}
\alpha\left(s\right)=\frac{1}{s}\frac{s^{2}+\left(2N-2\right)g_{0}^{2}}{s^{2}+\left(4N-2\right)g_{0}^{2}}.\end{equation}
 These expressions have only poles, but no branch points: Two  poles for $\beta_{k}\left(s\right)$, $s=\pm i2g_{0}\sqrt{N-\frac{1}{2}}$,
and three for $\alpha\left(s\right)$, $s=0,\pm i2g_{0}\sqrt{N-\frac{1}{2}}$,  see Fig.
\ref{fig:analytic_structure}. The
inverse Laplace transform is given by residues only, $\alpha\left(t\right)=\alpha_{p}\left(t\right)$
and $\beta_{k}\left(t\right)=\beta_{k}^{p}\left(t\right)$, \begin{equation}
\beta_{k}^{p}\left(t\right)=\frac{i\sin\left(2g_{0}\sqrt{N-\frac{1}{2}}t\right)}{\sqrt{2N-1}}\end{equation}
 and\begin{equation}
\alpha_{p}\left(t\right)=\frac{N-1}{2N-1}+\frac{N}{2N-1}\cos\left(2g_{0}\sqrt{N-\frac{1}{2}}t\right).\end{equation}
Expanding the above expression in $1/N$, we obtain the corrections $\delta\alpha\left(0\right)$ and
$\delta\beta_{k}\left(0\right)$ in Eqs. (\ref{eq:ap_t}, \ref{eq:bck_t}).

To calculate the $1/N$-correction when $\xi=g_{0}$, i.e. for maximum
inhomogeneity, we expand Eq. (\ref{eq:sum_even_couplings}) up to
the second order in $1/N$  at the poles, $s=\pm is_{0}$, $s_{0}=\sqrt{4N\left\langle g^{2}\right\rangle }$,
and also account for the second order corrections that come from the
positions of the poles, $s_{0}=\sqrt{4N\left\langle g^{2}\right\rangle -\frac{6}{5}g_{0}^{2}}$.

Performing this procedure we write the residues of $\alpha\left(s\right)$
at $s=\pm is_{0}$ as \begin{equation}
\textrm{Res}_{s=\pm is_{0}}\alpha\left(s\right)e^{st}=\frac{1}{\pm is_{0}}\frac{N_{0}+\delta N}{D_{0}+\delta D}e^{\pm is_{0}t},\end{equation}
 where 
  \begin{equation}
D_{0}=-\frac{2\left(\pm is_{0}\right)Ng_{0}^{2}}{\left(-s_{0}^{2}+N\left\langle g^{2}\right\rangle \right)^{2}},\label{eq:D0_N0}\end{equation}
 \begin{equation}
N_{0}=1+\frac{Ng_{0}^{2}}{3\left(-s_{0}^{2}+N\left\langle g^{2}\right\rangle \right)}\end{equation}
 are the  denominator and numerator obtained using just $s_{0}=\sqrt{4N\left\langle g^{2}\right\rangle }$. 
 Further,
\begin{equation}
\delta D=-\frac{2\left(\pm is_{0}\right)Ng_{0}^{2}}{\left(-s_{0}^{2}+N\left\langle g^{2}\right\rangle \right)^{2}}\frac{12g_{0}^{2}}{5\left(-s_{0}^{2}+N\left\langle g^{2}\right\rangle \right)},\label{eq:dD}\end{equation}
 \begin{equation}
\delta N=\frac{2Ng_{0}^{4}}{5\left(-s_{0}^{2}+N\left\langle g^{2}\right\rangle \right)^{2}}.\end{equation}
 are the first $1/N$-corrections. The average $\left\langle g^{2}\right\rangle =g_{0}^{2}/3$
is evaluated using $Q_{1}\left(g\right)$ with $\xi=g_{0}$.

Performing the summation over the two poles $\pm is_{0}$ we get 
\begin{equation}
\sum_{\pm is_{0}}\textrm{Res}_{s=\pm is_{0}}\alpha\left(s\right)e^{st}=\frac{2}{is_{0}}\frac{N_{0}+\delta N}{D_{0}+\delta D}\cos\left(s_{0}t\right),\end{equation}
 and expand it in the small parameter as\begin{multline}
\sum_{\pm is_{0}}\textrm{Res}_{s=\pm is_{0}}\alpha\left(s\right)e^{st}\\
=\frac{2}{is_{0}}\frac{N_{0}}{D_{0}}\left(1+\frac{\delta N}{N_{0}}\right)\left(1-\frac{\delta D}{D_{0}}\right)\cos\left(s_{0}t\right),\end{multline}
 where the first two terms in the product still have to be expanded
in the small correction to $s_{0}=\sqrt{4N\left\langle g^{2}\right\rangle }$,
and the last two have to be calculated using only the leading term
$s_{0}=\sqrt{4N\left\langle g^{2}\right\rangle }$.

The first two terms, which have to be calculated with $s_{0}=\sqrt{4N\left\langle g^{2}\right\rangle -\frac{6}{5}g_{0}^{2}}$,
are\begin{widetext}\begin{eqnarray}
\frac{2}{is_{0}}\frac{N_{0}}{D_{0}} & = & \frac{2}{i\sqrt{4N\left\langle g^{2}\right\rangle }}\left(1+\frac{3g_{0}^{2}}{20N\left\langle g^{2}\right\rangle }\right)\left(\frac{2}{3}-\frac{2}{5N}\right)\frac{-g_{0}^{2}N}{i2\sqrt{4\left\langle g^{2}\right\rangle }}\left(1-\frac{3}{5N}\right)\left(1+\frac{9}{20N}\right)\nonumber \\
 & \approx & \frac{1}{2}\left(1+\frac{9}{10N}\right)\left(1-\frac{6}{5N}\right)\approx\frac{1}{2}\left(1-\frac{3}{10N}\right)\end{eqnarray}
 \end{widetext} and the last two, which have to be calculated with
$s_{0}=\sqrt{4N\left\langle g^{2}\right\rangle }$, are \begin{equation}
\left(1+\frac{\delta N}{N_{0}}\right)\left(1-\frac{\delta D}{D_{0}}\right)=\left(1+\frac{3}{5N}\right)\left(1+\frac{12}{5N}\right)\approx1+\frac{3}{N}.\end{equation}
 Finally, the contribution from the poles $\pm is_{0}$ is\begin{equation}
\sum_{\pm is_{0}}\textrm{Res}_{s=\pm is_{0}}\alpha e^{st}=\frac{1}{2}\cos\left(s_{0}t\right)+\frac{27}{20N}\cos\left(s_{0}t\right).\label{eq:correction_s0}\end{equation}

The residue of $\alpha\left(s\right)$ at $s=0$ is also expanded
in the small corrections, \begin{equation}
\textrm{Res}_{s=0}\alpha\left(s\right)=\frac{N_{0}+\delta N}{D_{0}+\delta D}\approx\frac{N_{0}}{D_{0}}\left(1+\frac{\delta N}{N_{0}}\right)\left(1-\frac{\delta D}{D_{0}}\right),\end{equation}
 where \begin{equation}
\frac{\delta N}{N_{0}}=\frac{9}{5N},\quad\frac{\delta D}{D_{0}}=\frac{36}{5N}.\end{equation}
 And the contribution from the pole $s=0$ is \begin{equation}
\textrm{Res}_{s=0}\alpha\left(s\right)=\frac{1}{2}-\frac{27}{20N}.\label{eq:correction_0}\end{equation}
 The sum of Eqs. (\ref{eq:correction_s0}, \ref{eq:correction_0})
gives the correction $\delta\alpha\left(g_{0}\right)$ from Eq. (\ref{eq:ap_t}).

Then, we calculate the corrections to $\beta_{k}^{p}\left(t\right)=\textrm{Res}_{s=\pm is_{0}}\beta_{k}\left(s\right)e^{st}$.
Expanding the residues at $s=\pm is_{0}$,
 where $s_{0}=\sqrt{4N\left\langle g^{2}\right\rangle -\frac{6}{5}g_{0}^{2}}$,
we get\begin{eqnarray}
\textrm{Res}_{s=\pm is_{0}}\beta_{k}\left(s\right)e^{st} & = & \frac{-i\sqrt{2}g_{k}e^{\pm is_{0}t}}{s_{0}^{2}-N\left\langle g^{2}\right\rangle +2g_{k}^{2}}\frac{1}{D_{0}+\delta D}\\
 & \approx & \frac{-i\sqrt{2}g_{k}e^{\pm is_{0}t}}{s_{0}^{2}-N\left\langle g^{2}\right\rangle +2g_{k}^{2}}\frac{1}{D_{0}}\left(1-\frac{\delta D}{D_{0}}\right), \nonumber \end{eqnarray}
 where $D_{0}$ and $\delta D$ have already been calculated, see
Eqs. (\ref{eq:D0_N0}, \ref{eq:dD}). Similarly to the calculation
of $\alpha\left(s\right)$ we again expand and get\begin{widetext}
\begin{eqnarray}
\textrm{Res}_{s=\pm is_{0}}\beta_{k}\left(s\right)e^{st} & = & \frac{-i\sqrt{2}g_{k}}{Ng_{0}^{2}}\left(1-\frac{2\left(g_{k}/g_{0}\right)^{2}}{N}+\frac{6}{5N}\right)\frac{Ng_{0}^{2}}{\left(\pm i\right)4\sqrt{N\left\langle g^{2}\right\rangle }}\left(1-\frac{3}{5N}\right)\left(1+\frac{12}{5N}\right)e^{\pm is_{0}t}.\end{eqnarray}
 The sum over these residues, \begin{eqnarray}
\beta_{k}^{p}\left(g_{0}\right) & = & \frac{ig_{k}\sqrt{2}}{2\sqrt{N\left\langle g^{2}\right\rangle }}\left(1-\frac{2\left(g_{k}/g_{0}\right)^{2}}{N}+\frac{6}{5N}\right)\left(1+\frac{9}{5N}\right)\sin\left(s_{0}t\right)\nonumber \\
 & \approx & \frac{ig_{k}}{\sqrt{2N\left\langle g^{2}\right\rangle }}\left(1-\frac{2\left(g_{k}/g_{0}\right)^{2}}{N}+\frac{3}{N}\right)\sin\left(s_{0}t\right),\end{eqnarray}
 \end{widetext} is the correction from Eq. (\ref{eq:bck_t}) for
$\xi=g_{0}$.


\begin{thebibliography}{24}
\bibitem{diffusive_systems} In a diffusive system the Ehrenfest time is the time scale at which a minimal wave packet spreads over a characteristic spatial scale \cite{AleinerLarkin,AdagideliBeenakker, RahavBrouwer}.
\end{thebibliography}
\end{document}